\documentclass[%
reprint,
 amsmath,amssymb,
 aps,
]{revtex4-2}

\usepackage{graphicx}% Include figure files
\usepackage{dcolumn}% Align table columns on decimal point
\usepackage{bm}% bold math
\usepackage{subfigure} % for subfigures
\usepackage{subcaption}
\begin{document}
\preprint{APS/123-QED}
\title{Order, Collectivity,  Structured Yrast Lines, and Correlated Energies in Random Bosonic Systems}
\author{D. Mulhall}
\address{Department of Physics/Engineering,\\
 University of Scranton,\\ Scranton, \\Pennsylvania 18510-4642, USA}
\begin{abstract}
Signatures of order and collectivity appear in a broad range of one and two level boson systems. Toy systems of $N$ particles on a single $j=3,\,4\dots 7$ level, and on 2 levels $(j_1,\,j_2)$  with $j_1+j_2 \leq 6$ were studied. Surprising novel features in the yrast lines were consistent with boson condensates. Strong signatures of order and collectivity were the norm. These included the usual energy ratios and quadropole transition strengths for vibrational and rotational bands. There were pronounced correlations in energy ratios, sharp peaks in the Alaga ratios and fractional collectivity across all the 1-level systems, as well as signatures of triaxiality. Yrast lines consistent with a deformed rotating core were common.
\end{abstract}
\maketitle

%\tableofcontents

\section{Introduction}

It is nearly 25 years since the publication of the surprising result that random interactions, constrained only by global symmetries, yield spectra with the signatures of regular interactions \cite{jbd1}. Order and collectivity in nuclei at low energies is inferred from simple robust spectral correlations. Examples of such inferences include pairing, rotational and vibrational bands, and collectivity. That these signatures of regularity can be generated with the  Two Body Random Ensemble (TBRE) in systems of fermions or bosons \cite{bijker99, *bijker00a, *bijker00b, *bijker02b}, raises some serious questions. See \cite{zelevinsky04,*Zhao04} for a comprehensive review.

In the original paper \cite{jbd1} Johnson et al. showed that an ensemble of random two-body interactions that conserve angular momentum, have a proportion, $f_0$, of spin-0 ground states that is huge in comparison to the proportion of spin-0 states in the Hilbert space. They also reported parabolic yrast lines, what looked like a pairing gap, signatures of the generalized seniority model, and other evidence of collectivity.   Attempts were made to find explanations for these regularities based on the details of the models that produced them \cite{zhao01,*zhao02b,*zhao02c}.  This included a series of studies of the IBM with random interactions by Bijker et al \cite{bijker99,bijker00a, bijker00b}. They saw the by now familiar large $f_0$, and evidence of rotational and vibrational bands in the distinct peaks in $R_{42}$ (defined below). Furthermore, they saw a pronounced excess probability $f_{\textrm{Jmax}}$ that the ground state has maximum spin. These results were also seen in the   vibron model  \cite{bijker02}, where $N$ bosons with spin 0 and 1 interact randomly.  In \cite{zhao02b} Zhao et al. showed collective vibrations in a truncated Hamiltonian were made of $sd$ pairs, while the collective rotations were not. A comparison by Horoi et al. \cite{horoi01} of random vs realistic interactions in the shell model supported a picture of these random spectra looking realistic, but the corresponding wave functions having a small overlap with their realistic counterparts. In \cite{johnson07} Johnson and Nam saw that in the random IBM and the nuclear shell model there are strong correlations between  the  energy ratios $R_{82},\,R_{62}$ and $R_{42}$, where $R_{I2}=(E(I_1)-E(0_1))/(E(2_1)-E(0_1)$ etc.  $E(I_1)$ is the lowest energy with angular momentum $I$, this is called the yrast energy. These correlations were seen by Lei et al. \cite{lei11}  in systems of $sd$-bosons.

The work on the TBRE in the IBM and vibron model  showed signatures of collectivity.  In \cite{Lu14} Lu et al. saw strong correlations in the TBRE on the IBM between $R_{42}$ and both $R_{62}$ and $R_{82}$. They explained this in terms of $d-$boson condensates.  The question of what values these collective observables take for a broader range of $(j_1,j_2)$ is interesting.

This work concerns realizations of the TBRE on toy systems of $N$  bosons on 1 and 2 levels. For both types of system  the distribution of ground state spins $J_\textrm{{gs}}$ and the energy ratio of yrast states was examined,  (an yrast state is the lowest energy $J$-state). The reduced transition ratios $B(E2,4_1\rightarrow 2_1)$ and $B(E2,2_1\rightarrow 0_1)$, as well as their ratio $B_{42}$, were calculated. Separate calculations were made for a series of $(N,j)$ systems. These were the correlations between the quadropole moment  of  the lowest $J=2$ states, $\mathcal{Q}(2_1)$  and $\mathcal{Q}(2_2)$, the fractional collectivity, and measures of triaxiality, as well as the Alaga ratio (which is another measure of collectivity). The main results were that ALL of these systems had strong indicators of collective behavior. Furthermore, there were novel structures in the yrast lines in these systems. The 1-level systems had quadratic yrast lines built on a deformed ground state core.

Regarding notation,  lower case $j,m$ is used for single particle angular momentum and projection,
and upper case $J,M$ is used for many-body angular momentum. The terms ``angular momentum" and ``spin"
are used interchangeably. The maximum value of the total spin is $J_{\textrm{max}}$, and the  ground state spin
is $J_{\textrm{gs}}$.  The probability that
$J_{\textrm{gs}} = J$ is $p(J)$. Specific cases have their own symbol, $f_0 =p(0)$, the
probability of a spin-0 ground state (these states will also be referred to as
$0_{\textrm{gs}}$). Similarly, $f_{\textrm{max}}=p(J_{\textrm{max}})$. The probability that
the two lowest energies have $J=0,\,2$ is written $f_{0-2}$. The $n$-th level with spin-$J$ is labeled
$J_n$, making the yrast energies $E(J_1)$.

The next section has the details of the Hamiltonian and various single particle operators. Section \ref{gsj}  concentrates on the distribution of ground state spins and the yrast lines. Section \ref{1level} will deal with the measures of collectivity in the 1-level systems, specificallydescribing the distribution of energy ratios, $R_{42}$, transition rate ratios, $B_{42}$, triaxiality, fractional collectivity, Alaga ratios, and correlations between $\mathcal{Q}(2_1)$ and $\mathcal{Q}(2_2)$. Section \ref{2level} will describe correlations in the yrast energies of 1 and 2-level systems.

\section{The Hamiltonian \label{ham}}
This work explores a toy model of $N$ bosons having random 2-body interactions that conserve angular momentum.
Models for both 1 and 2 levels were constructed. The 1-level Hamiltonian is
\begin{equation}
H=\sum_{L\Lambda}V_{L}P^{\dagger}_{L\Lambda}P_{L\Lambda},  \label{2.1}
\end{equation}
where the operators for a pair of spin-$j_1$ bosons coupled to total angular momentum  $L$ and
projection $\Lambda$ are defined as
\begin{eqnarray}
P^{\dagger}_{L\Lambda}=\frac{1}{\sqrt{2}}\sum_{m_i m_k}C^{L\Lambda}_{j_1 m_i j_1 m_k}
a^{\dagger}_{j_1 m_i}a^{\dagger}_{j_1 m_k} , \\
 P_{L\Lambda}=\frac{1}{\sqrt{2}}
\sum_{m_i m_k}C^{L\Lambda}_{j_1 m_i j_1 m_k}a_{j_1 m_k}a_{j_1 m_i}.
\label{1a}
\end{eqnarray}
 $L$ can have any even value, $L=0,2,4\dots 2j_1$. $C^{L\Lambda}_{j_1 m_i j_1 m_k}$ are the
 Clebsch-Gordan coefficients $\langle L\Lambda|j_1 m_i;j_1 m_k \rangle$ , and the two-particle states
 $|2;L\Lambda\rangle=P^{\dagger}_{L\Lambda}|0\rangle$ are properly normalized. The interaction
 parameters are $V_{L}$ and these numbers define our ensemble. They are normally distributed with
 $(\mu,\sigma)=(0,1)$. In practice they are taken from a table with $5\times 10^4$ rows, meaning that
 the $i^{\textrm{th}}$ spectrum of each $(N,j_1)$ ensemble have the same ${V_L}$.

 The 2-level systems has 8 terms. There is a single particle term for each level, $ \sum_{i=1,2}
 \epsilon_i a^{\dagger}_i a_i $ which added nothing to the dynamics, but is there to break any extra
 degeneracies. The single particle energies $\epsilon_i$ were $\pm 0.001$. Each 2-body interaction
 could move 0,1 or 2 particles from either level. Writing the interaction parameters  $V_{Lijkl}$, to
 denote the initial and final levels of each of the particles, and writing $P_{Lij}$ as the operator
 that creates a pair of particles, with  one on level $i$ and the other on level  $j$, gives
 \begin{eqnarray}\label{ham2}
    H = &&\sum_{i=1,2} \epsilon_i a^{\dagger}_i a_i  \\
    + &&\sum_{L\Lambda}V_{L1111}P^{\dagger}_{L\Lambda 11}P_{L\Lambda 11}\nonumber\\
    +   &&\sum_{L\Lambda}V_{L2222} P^{\dagger}_{L\Lambda 22}P_{L\Lambda 22} \nonumber \\
    +  &&\sum_{L\Lambda}V_{L1212} P^{\dagger}_{L\Lambda 12}P_{L\Lambda 12}  \nonumber\\
    + 1/2 &&\sum_{L\Lambda}V_{L1112}(P^{\dagger}_{L\Lambda 11}P_{L\Lambda 12} +P^{\dagger}_{L\Lambda
    12}P_{L\Lambda 11}) \nonumber\\
    + 1/2 &&\sum_{L\Lambda}V_{L1122}(P^{\dagger}_{L\Lambda 11}P_{L\Lambda 22} +P^{\dagger}_{L\Lambda
    22}P_{L\Lambda 11}) \nonumber\\
    + 1/2 &&\sum_{L\Lambda}V_{L1222}(P^{\dagger}_{L\Lambda 12}P_{L\Lambda 22} +P^{\dagger}_{L\Lambda
    22}P_{L\Lambda 12}) \nonumber
 \end{eqnarray}

The Hamiltonians were diagonalized in the $M=0$ subspace, and the $\textbf{J}^2$ operator was used to
assign   $J$-labels.
The reduced transition rate, or $B(E2)$ values, and the intrinsic quadropole moment of a state are
calculated in terms of $\mathcal{M}_{2\mu}$, the multipole operator  \cite{poan}:
\begin{eqnarray}\label{beq}
    B(E2,J_i\rightarrow J_f)&=&\frac{1}{2J_i + 1} \sum_{\mu M_i M_f} |\langle J_f M_f |
    \mathcal{M}_{2\mu}|J_i M_i \rangle | ^2 \nonumber \\
    \mathcal{Q}(J)&=&\frac{\langle J J | \mathcal{M}_{2 0}|J J \rangle}{C^{J J}_{2 0,J J} C^{J 0}_{2
    0,J 0}}
\end{eqnarray}

 On one level,
 \begin{equation}
    \mathcal{M}_{2\mu}= (-1)^{-\mu}  \sum_{m_i m_k} (-1)^{j-m_k}C^{2  -\mu}_{j m_i,j -m_k}\, a^\dag_{j
    m_i}\, a_{j m_k}.
\end{equation}
On two levels there are cross terms. In general
 \begin{equation}\label{m2}
 \mathcal{M}_{2\mu}= \alpha \mathcal{M}^{11}_{2\mu}+\beta \mathcal{M}^{22}_{2\mu}+\frac{1}{2}\, \gamma
 (\mathcal{M}^{12}_{2\mu}+\mathcal{M}^{21}_{2\mu})
\end{equation}
where
 \begin{equation}
    \mathcal{M}^{\alpha \beta}_{2\mu}= (-1)^{-\mu}  \sum_{m_i m_k} (-1)^{j_\alpha-m_k}C^{2
    -\mu}_{j_\alpha m_i,j_\beta -m_k}\, a^\dag_{j_\alpha m_i}\, a_{j_\beta m_k}.
\end{equation}
In the IBM, where $(j_1,j_2)=(0,2)$, $\alpha=0,\,\beta = -\sqrt{7}/2$, and $\gamma = 1$
\cite{bijker00b}. In this work both  $\chi=0$ and 1 are used in the 2-level calculations for $\mathcal{Q}(J)$ and $B(E2,J_i\rightarrow J_f)$.
used

\section{The ground state spins \label{gsj}}
\begin{figure}
  \includegraphics[width=\linewidth]{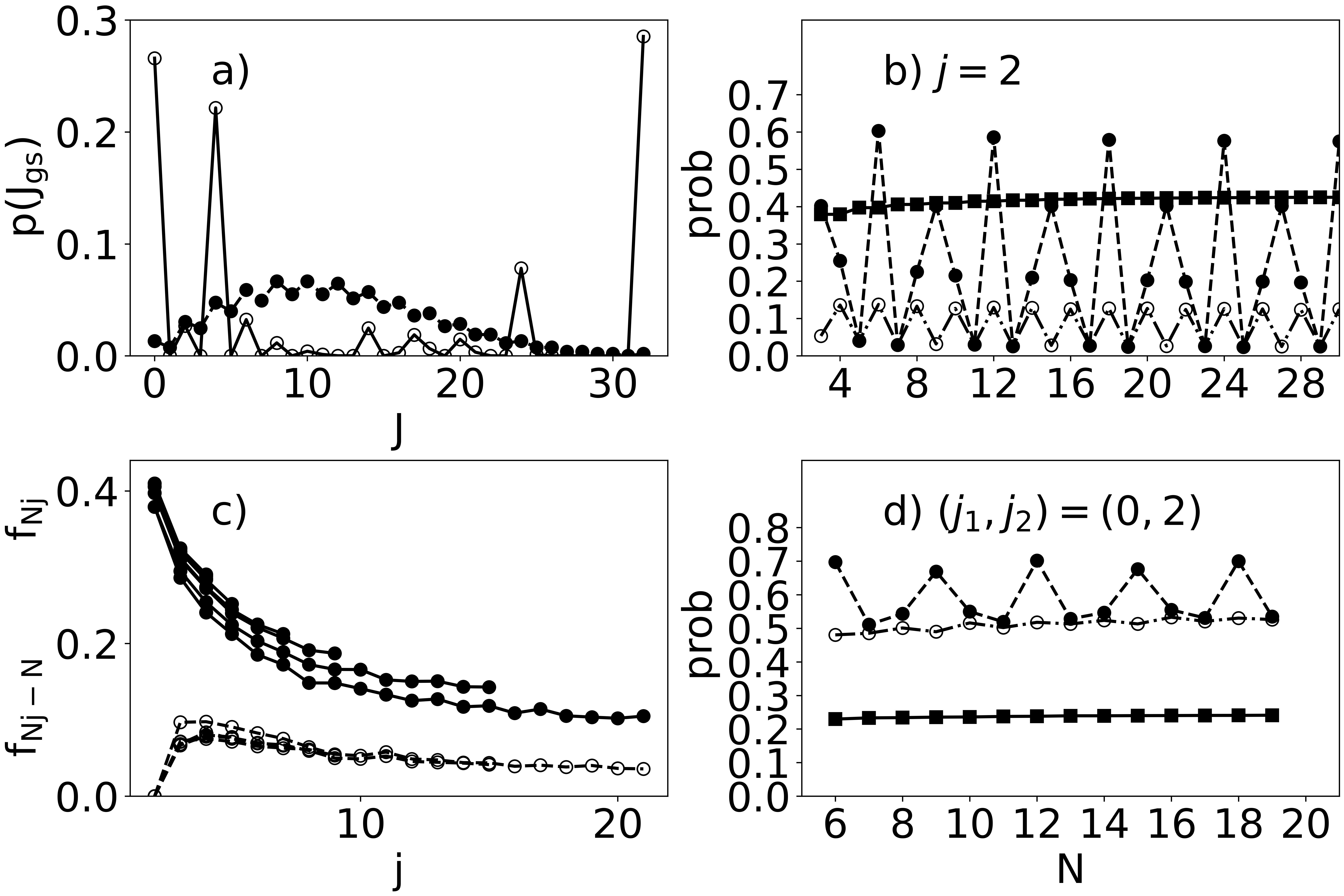}
  \caption{  a)   $p(J)$ of the ground state spin for $(N,j)=(8,4)$. The dashed line (solid circles) shows  the degeneracy of $J$ normalized to be a probability density. The solid line (open circles) is the ensemble result. Notice the peaks in $p(J)$ at $J=0,\,4,\,24,$  and $32$. b) The probabilities $f_0$(dashed, solid circles), $f_{0-2}$(dash-dot, open circles) and $f_{\textrm{Jmax}}$(solid line, squares) as a function of particle number for the single level $j=2$ systems. c) $f_{Nj}$ (solid line, solid circles)  and $f_{Nj-N}$ (dashed line, open circles) vs $j$ for various values of $N$. The longer lines correspond to smaller $N-$values. The value of $f_{Nj}$ increases with increasing $N$ for a fixed value of $j$. d) same as b) but for the two level $(j_1,j_2)=(0,2)$ systems. Notice that compared to b),$f_{\textrm{Jmax}}$ is lower and $f_{0-2}$ is higher.}\label{fig.pjd}
\end{figure}

One of the most striking findings of earlier studies was that the distribution of ground state spins,
$p(J)$, does not follow the distribution of allowed angular momenta, $d_J$ for random $J$-conserving
interactions. High values of $f_0$ were seen in fermionic  \cite{jbd1} and bosonic  \cite{bijker99} systems, and this question was attacked vigorously  early on \cite{zelevinsky04,*Zhao04}. In Fig.~\ref{fig.pjd} a) we see  $p(J)$  for  $(N,j) = (8,4)$. This system is representative of the even-$N$ systems studied. Also plotted is a probability function made by rescaling $d_J$. The graphs are dramatically different, with $p(J)$ having pronounced and consistent peaks at 4 special values of $J$, namely $J=0,\,j,\,Nj-N$ and $Nj$. We call the probabilities of these values  $f_0,\,f_\textrm{j},\,f_{\textrm{Jmax-N}}$ and $f_{\textrm{Jmax}}$. There is a high probability that the ground state has $J_{\textrm{gs}}=0$ for even $N$. Similarly for odd $N$, $f_\textrm{j}$ is pronounced, and is consistent with a single valence spin-$j$ particle on an even-$N$ spin-0 core. In Fig.~\ref{fig.pjd} c) the plots for $f_{\textrm{Jmax}}$ and $f_{\textrm{Jmax-N}}$ show a very weak dependance on $N$, and on $j$ above $j=7$.  This was seen in \cite{zhao03}.  The fraction of spectra with $J_{\textrm{gs}}=J_{\textrm{max}}$ and $J_{\textrm{max}}-N$  together account for around 20\% of total spectra.  This is  remarkable given that the degeneracy of $J_{\textrm{max}}=Nj$ is unity for all choices of $(N,j)$.  The trends for $f_0$ and $f_{Jmax}$ carry over to the 2-level systems, see Table \ref{tab:table1}. It is important to note that the special case of $d-$bosons ($j=2$) can be solved analytically, see \cite{White23}.

\section{The  yrast lines \label{yrast}}
\begin{figure}
  \includegraphics[width=\linewidth]{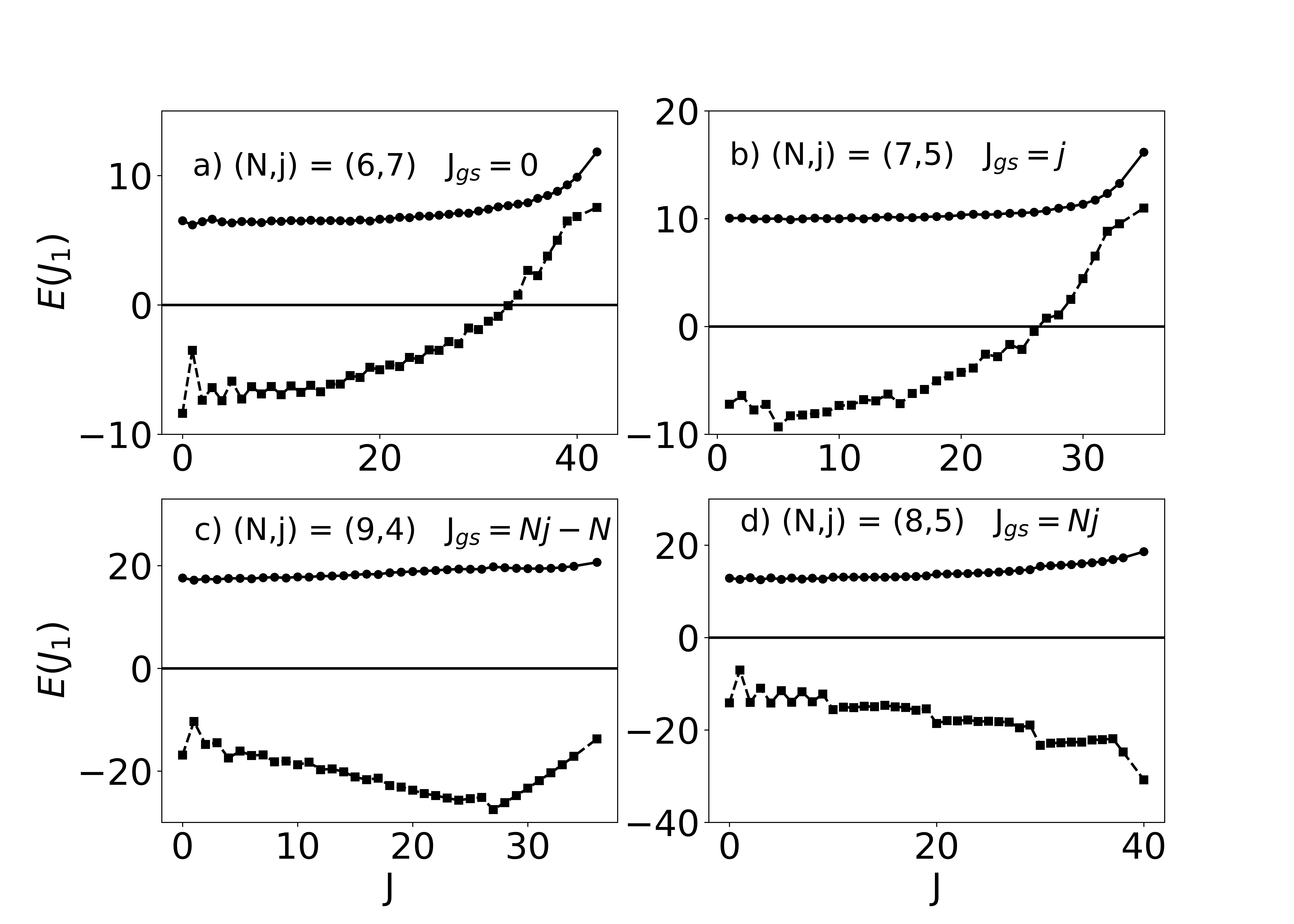}
   \caption{ Representative yrast lines $E(J_1)$ (lower curves, squares), and their associated standard deviations    (upper curves, circles), for the four most common cases. (a) $J_{\textrm{gs}}=0$ with
   $(N,j)=(6,7)$.  (b) $J_{\textrm{gs}}=j$ with $(N,j)=(7,5)$.  (c) The lines for $J_{\textrm{gs}}=Nj-N$ had a very
   distinct shape. Here we have $(N,j)=(8,5)$. The lines always consist of a  decreasing section
   finishing with an abrupt change at $J=Nj-2j$. The last section is always from $Nj-N-1$ to $Nj$.
   Though it looks linear, it is not. See Fig.~\ref{fig.yrastNjmN}.  (d) The $J_{\textrm{gs}}=Nj$ case, with
   $(N,j)=(9,4)$. These lines all followed the same pattern, with no exception. The lines were flat
   steps, of width $2j$, and a half step of width $j$  beginning at $J=0$ when $N$ is odd. }\label{fig.yrastESigma}
\end{figure}

The ensemble average  yrast lines, and corresponding standard deviations, were calculated for each of
our $(N,j)$ ensembles. This was also done for subsets of spectra with
$J_{\textrm{\textrm{gs}}}=0,\,j,\,J_{\textrm{max}}-N$ and $J_{\textrm{max}}$. These yrast lines have
some very unexpected features.   Fig.~\ref{fig.yrastESigma} shows representative plots for each of
these cases. Each panel shows the average and standard deviation of $E(J_1)$. The standard deviations all
start flat and bent upwards a little at higher $J$. In each case, the main features of the lines are
robust, they don't depend on the values of $N$ and $j$, other than that the $J_{\textrm{gs}}=0$ or $j$
predominance is sensitive to whether $N$ is  even or odd. In each case the standard deviation of the
yrast energies is relatively flat, and the same magnitude as the low-$J$ yrast energies. Here is a
summary of the  the main results for the yrast lines for the 1-level systems.

\subsection{The  $J_{\textrm{\textrm{gs}}}=0$ or $j$ systems}
\begin{figure}
  \includegraphics[width=\linewidth]{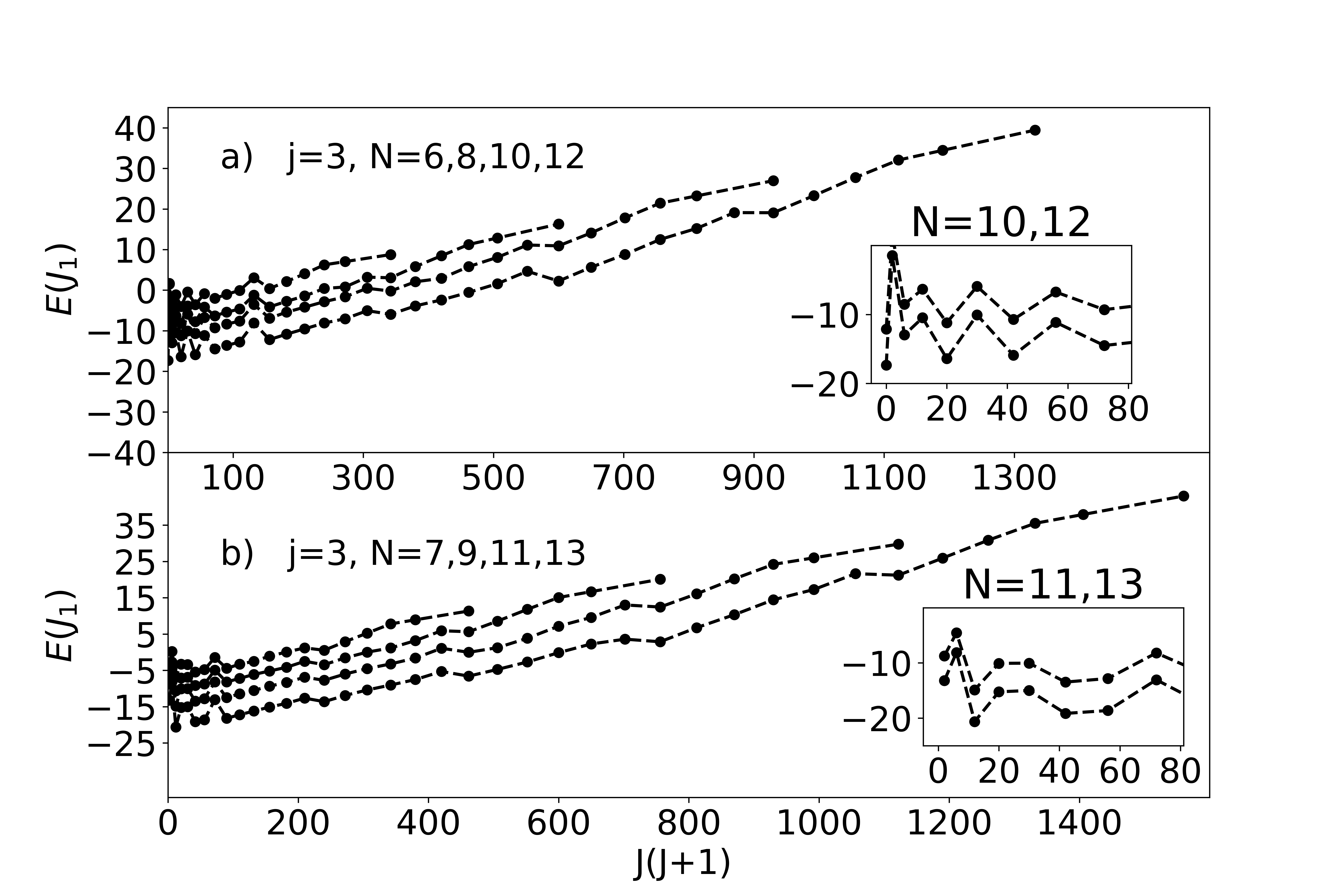}
   \caption{$E(J_1)$ vs $J(J+1)$ for the $(N,3)$ systems. $N$ ranges from 6 to 13. The insets show the
   first step for the high $N$ values.  a) Even $N$. Notice that the lines are piecewise linear, with
   slopes increasing with $J$. The steps in $J$ are from $0 \rightarrow 2j-1,\,2j \rightarrow
   4j-1,\dots 8j\rightarrow 10j-1$. (Note  $J=Nj-1$ is not allowed). b) the odd values of $N$.
   Because $J=0$ is forbidden when both $N$ and $j$ are odd, the first step is $1\rightarrow j-1=2$.
   }\label{fig.YrastLinesJgs0andj}
\end{figure}

The even-$N$ systems favor $J_{\textrm{gs}}=0$ while the odd-$N$ systems favor $J_{\textrm{gs}}=j$. This is consistent with a valence  particle sitting on a spin-0 core. The yrast lines for both these cases are similar and striking. When $E(J_1)$ is plotted against $J(J+1)$ the lines are piecewise linear with jumps at regular intervals of $2j$, see Fig.~\ref{fig.YrastLinesJgs0andj}. The odd-$N$ lines start off with a mean negative slope for $0 \leq j$, followed by a set of $(N-1)/2$ linear sections of length $2j$. Note $J=0$ is not allowed when both $N$ and $j$ are odd. The even-$N$ lines are composed of $N/2$ sections of length $2j$. The first section has a strong odd-even effect, and there is a large jump in energy from $J=0$ to 1,  $E(J_1)$ is linear in $J(J+1)$ in all but the first section. In the first section  $E(J_1)$ is linear in $J(J+1)$ for even-$J$. The slope of these sections is $1/(2\mathcal{I})$, where $\mathcal{I}$ is a dynamical moment of inertia. These slopes tend to increase as $J$ increases. A physical picture of a rigid deformed rotating core jumping to a different configuration each time $J$ changes by $2j$ is tempting. Why this configuration would have a lower $\mathcal{I}$ is puzzling.

\subsection{The  $J_{\textrm{gs}}=Nj-N$ systems} 
\begin{figure}
  \includegraphics[width=\linewidth]{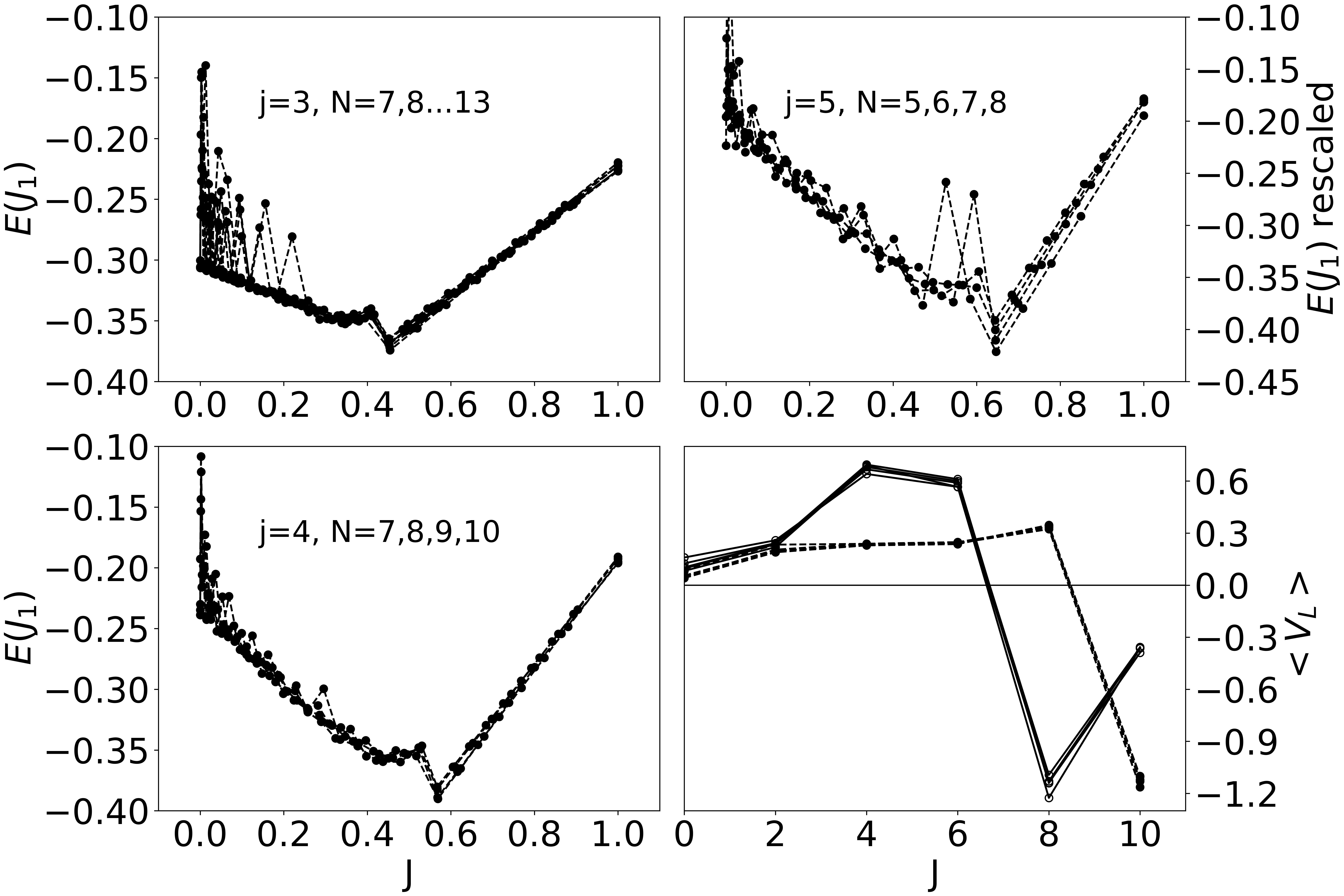}
   \caption{The  $(N,3),\,(N,4)$ and $(N,5)$ system  yrast lines  with $J_{\textrm{gs}}=Nj-N$. The lines have been rescaled as described in the    main text. Bottom right panel has the average interaction parameters $<V_L>$ for $j=5$ systems with    $N=4,5\dots8$ that resulted in $J_{\textrm{gs}}=Nj$ and $Nj-N$. There are two closely bunched sets of 5    lines, one for each $N$. The $J_{\textrm{gs}}=Nj-N$ spectra came from sets  $\{V_L\}$ with $V_8$ minimum (solid line, open circles). The $J_0=J_{\textrm{max}}$ spectra came from  $\{V_L\}$ where $V_{2j=10}$ is is negative, all others positive (dashed, solid circles).}\label{fig.yrastNjmN}
\end{figure}

This set accounts for around 8\% of the ensemble
across a range of $N$ and $j$, see Fig.~\ref{fig.pjd} (b). The corresponding yrast lines have a
consistent shape, all with a steady decrease in $E(J_1)$ leading to a small flat region, before an abrupt
drop to a positive steep section, see   Fig.~\ref{fig.yrastESigma} (c).  Rescaling the axis reveals the
true consistency  of these systems. In Fig.~\ref{fig.yrastNjmN} we see the yrast lines for a range of
$N$ with $j=3$, upper panel, and $j=4$, lower panel. The energy is rescaled as $E(J_1)/(N(N-1))$, and the
total angular momentum squared is rescaled as $J(J+1)/(Nj(Nj+1))$. The lines start with a negative
linear section, leading to a flat section, abruptly changing to a steep linear positive section. The
smaller odd $J$ values in the $j=3$ systems deviate from the trend, but in all three graphs the trend
is strong otherwise.

The distribution of interaction parameters that gives $J_{\textrm{gs}}=Nj-N$ is shown in
Fig.~\ref{fig.yrastNjmN}. Not surprisingly, $V_{2j-2}$ is the minimum. Naively one might expect the
ground state to be a condensate of $L=2j-2$ pairs, with an energy $E_{\textrm{gs}}=(1/2) N(N-1)V_{2j-2}$,
but this isn't the case, the real ground state energy is only around 2/3  of this value. The section $Nj-N \leq J < Nj$ is very linear. This is exactly what one would expect if the ground state consisted ofa rigid rotating deformed condensate of pairs of spin-$2j-2$.

\begin{figure}
  \includegraphics[width=0.9\linewidth]{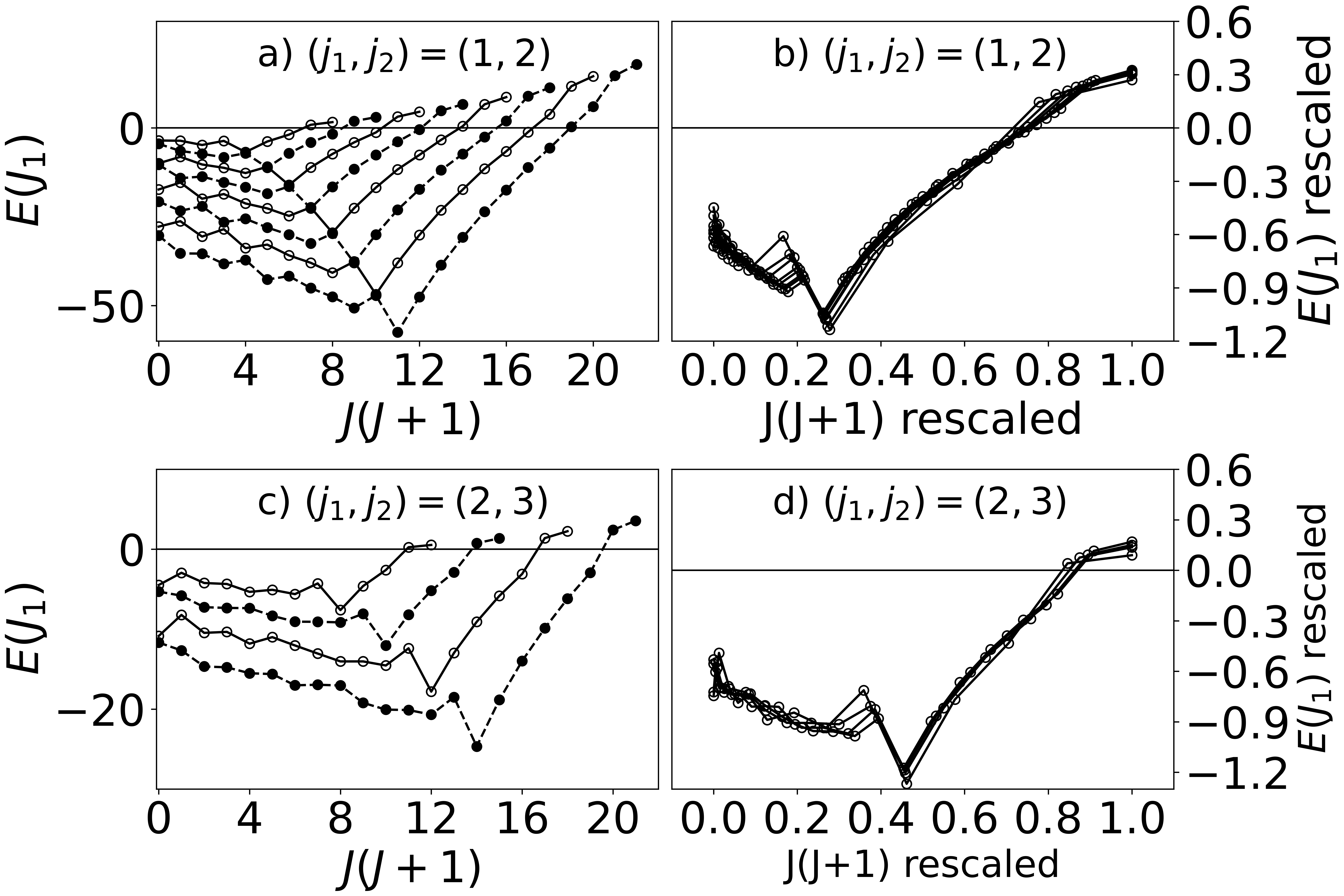}
   \caption{The yrast lines of the $J_{\textrm{gs}}=Nj_2-N$ subset for the $(N,1,2)$ and $(N,2,3)$
   systems. a) $N=4,5\dots 11$, with even-$N$ (dashed, solid circles), odd-$N$ (solid line, open circles). $(j_1,j_2)=(1,2)$. The    $E(J_{\textrm{gs}})$ of an $N$ particle system is close to the $E(J_{\textrm{gs}}-1)$ of the $N-1$    particle system. This leads to aliasing in the plot. b) The lines in a) rescaled as in Fig.4. c)    $N=4,5,5,7$, with even-$N$ (dashed, solid circles) and odd-$N$ (solid line, open circles). $(j_1,j_2)=(2,3)$. d) The lines in c) rescaled.
   }\label{fig.yrastNjmN2levels}
\end{figure}

In the selection of $(j_1,j_2)$ examined in this work, there were 5 that had $f_{\textrm{Jmax}-N}>5\%$.
They were $(0,3),(1,2),(1,3),(1,4),(2,3)$. The corresponding yrast lines are shown in Fig.~\ref{fig.yrastNjmN2levels} for the
$(1,2)$ and $(2,3)$ systems.  In the $(1,2)$ systems, we see a coincidence that leads to an aliasing
effect in the plot of the yrast lines for $N=4,5\dots 11$ systems. Write the ground state energy for
the $N$ particle system as $E_N(J_{\textrm{gs}})$. Note that, for $j=2$,  $J_{\textrm{gs}}=J_{\textrm{max}}-N=Nj-N=N$. We
see empirically in Fig 5 that $E_N(J_{\textrm{gs}})=E_{N_1}(J_{\textrm{gs}}-1)$, or $E_N(J=N)=E_{N+1}(J=N)$ leading to
the aliasing in Fig.~\ref{fig.yrastNjmN2levels}.

\subsection{The  $J_{\textrm{gs}}=J_{\textrm{max}}$ systems}  
\begin{figure}
  \includegraphics[width=\linewidth]{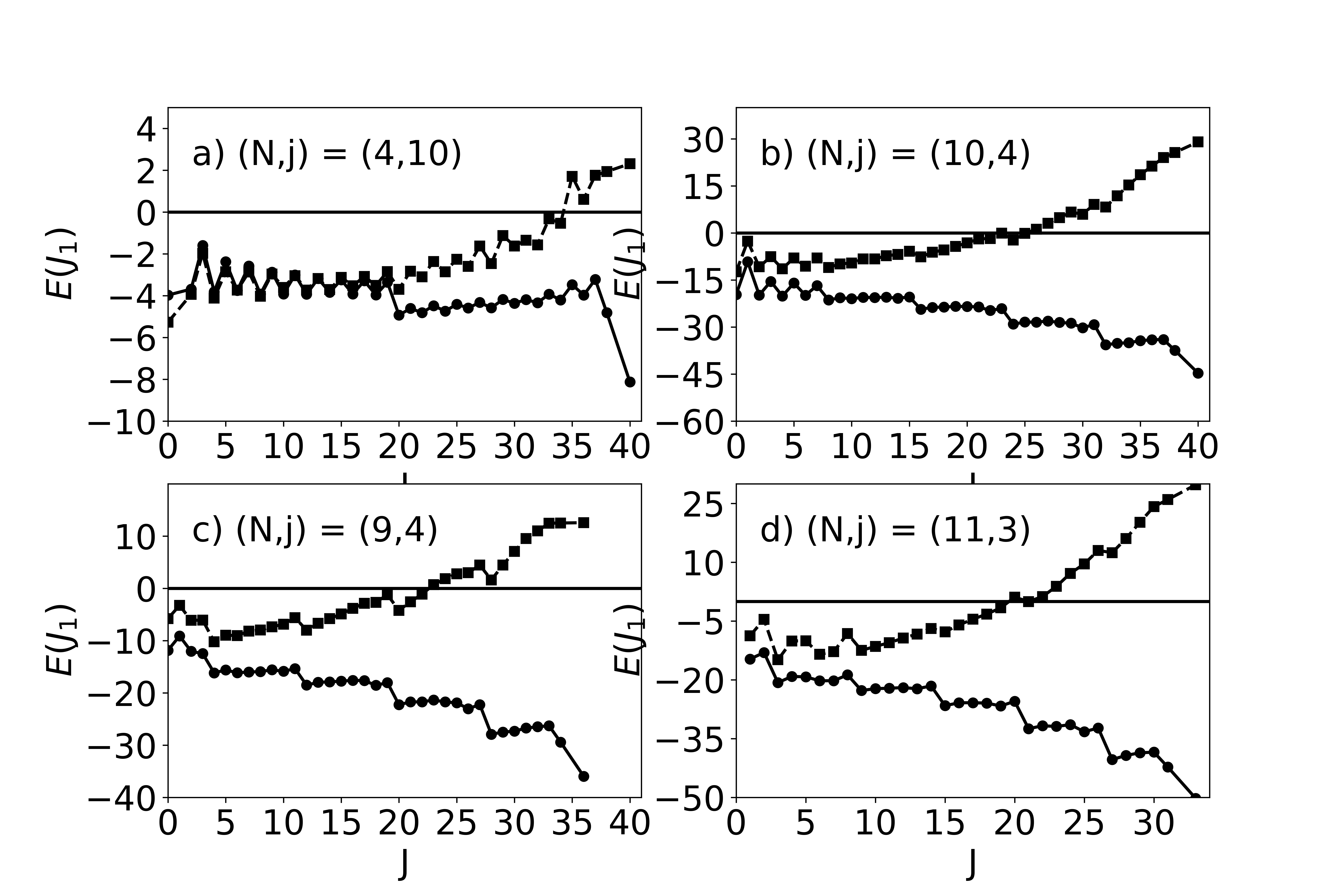}
   \caption{The yrast lines for  $J_{\textrm{gs}}=0$ (squares)and $J_{\textrm{gs}}=J_{\textrm{max}}$ (circles) for various $(N,j)$. The yrast lines    coincide to within a zero or constant offset for the region $0\leq J<2j-1$, and then diverge.  The    $J_{\textrm{gs}}=J_{\textrm{max}}$ lines clearly show $N/2$ steps of width $2j$. (a)  Because $(N,j)=(4,10)$ there are    2 steps of width 20 to the maximum $J=40$. (b) This time we have $(N,j)=(10,4)$, giving  5 steps of width 8.   In (c) we see    a kinks in the $J_{\textrm{gs}}=0$ yrast line at $J=j,\,3j,\,5j$ and $7j$. These kinks coincide with the   steps in the  $J_{\textrm{gs}}=J_{\textrm{max}}$ line. d) Again we see the yrast lines differ by a small offset for the first step. Also present are the kinks in the $J_{\textrm{gs}}=0$ line.  }\label{fig.yrastJ0JmaxSample}
   \end{figure}

\begin{figure}
  \includegraphics[width=\linewidth]{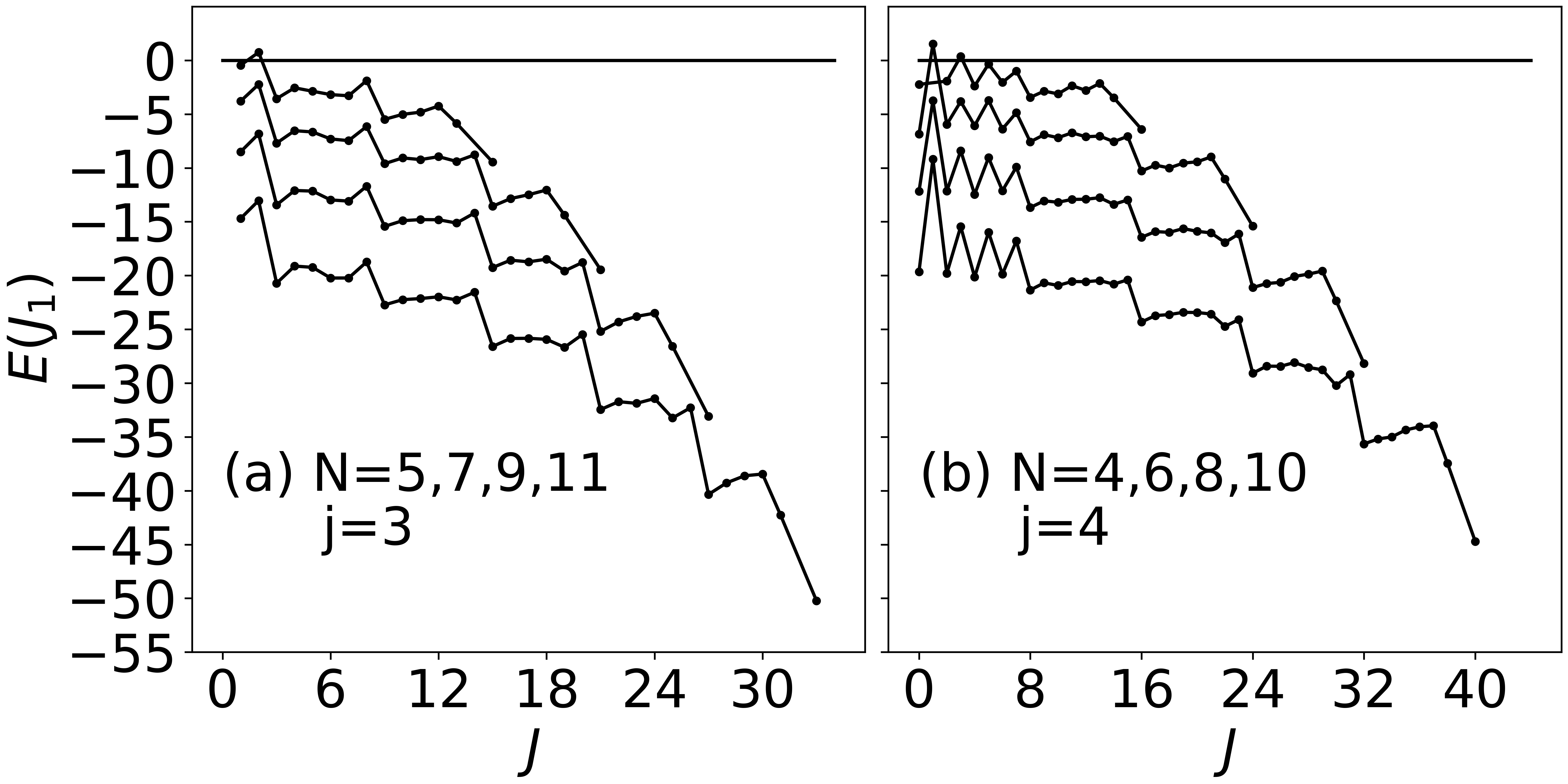}
   \caption{The yrast lines for $J_{\textrm{gs}}=J_{\textrm{max}}$. Values of $N$ are apparent from the endpoints, as $J=Nj$. The endpoint of the $N-2$ system coincides with the second-to-last step of the $N$ particle yrast line. (a) Odd-$N$ on $j=3$. There are $N$/2 steps The first step has only $J=1$ and 2, as $J=0$ is not allowed when both $N$ and $j$ are odd.. Notice, for example,  the endpoint of $N=5$ line, at $J=15$ coincides with the third energy step in the $N=7$ line, ending at $J=14$. (b) similar to (a) with $j=4$.    }\label{fig.yrastjmaxendpoints}
\end{figure}

These yrast lines are made of $N/2$ steps of length $2j$ when $N$ is even. When $N$ is odd, they start with a section of length $j$, followed by $(N-1)/2$ steps of length $2j$. See Fig.~\ref{fig.yrastJ0JmaxSample} and \ref{fig.yrastjmaxendpoints}. Furthermore, the first step of these lines matches to within a constant factor, the $J_{\textrm{gs}}=0$ lines for even-$N$, and the $J_{\textrm{gs}}=j$ lines for odd $N$, as well as the $J_{\textrm{gs}}=J_{\textrm{max}}-N$ for both odd and even $N$ .

The stepped structure of the $J_{\textrm{gs}}=J_{\textrm{max}}$ lines can be  understood as
condensate of non interacting spin-$2j$  quasiparticle pairs.   The average interaction parameters
$<V_L>$ that give rise to these lines is shown in Fig.~\ref{fig.yrastNjmN} (d). This distribution is
independent of $N$. The range of $L$ increases with $j$ ($L$ can have any even value from 0 to $2j$),
the shape of the distribution doesn't change. The distribution has all small positive parameters for
$L\leq 2j-2$ and large negative value for $V_{2j}$. This immediately suggests that the lowest energies
will be condensates of pairs with spin $2j$. Consider the system $(N,j)=(10,4)$ as an example, see
Fig.~\ref{fig.yrastJ0JmaxSample}. The maximum $J$ is 40. The yrast energy for the maximum angular momentum is
$E(J_{\textrm{max}})=(1/2) N(N-1) V_{2j}$, or $45V_8$. This state is unique, and each particle is part of
an $L=2j$ pair. As soon as we get to $J<40$ there is a pair broken, the system looks like a system of 4
spin-8 quasiparticles with their spins aligned to $J=32$. The 2 valence particles make pairs of spin
0,2,4,6.  As  one particle is not bonded to the others  the energy is now $(1/2) (N-1)(N-2)V_{2j}$,
or $36 V_8$. The states with  $32\leq J<40$ are made on this stretched $J=32$ core. When another pair
is broken, there is one less particle involved in making these pairs, and the yrast energies for
$24\leq J<32$ are constant at close to $(1/2) (N-2)(N-3) V_8=28V_8$. The actual energies are higher than
the pure $V_8$ contribution, but this is to be expected because the other interaction parameters are
small and positive. The ground state energy of
an $(N-2,j)$ system is the same as the last step of the $(N,j)$ system, which is consistent with this
picture. See Fig.~\ref{fig.yrastjmaxendpoints}.

The structure of the yrast line suggests a ferromagnetic alignment of these spin-$2j$ quasiparticles
coupled to the residual angular momentum of the unpaired particles. This residual angular momentum goes
from 0 to $2j-1$. If it went to $2j$ then that would constitute another spin-$2j$ quasiparticle
(putting the system on the next step).

\section{Collectivity and Shapes in the 1-level systems. \label{1level}}
Ensembles for a selection of 1-level systems were made to search for signatures of collectivity. The systems were $(N,j)=(6,6),\,(6,7),\,(8,4),\,(8,5),\,(10,4),(12,3)$, which had $f_{0-2}=(0.06, 0.07, 0.14, 0.12, 0.12, 0.20)$ respectively. Each ensemble had $2\times 10^4$ spectra.  The 0-2 subset, those $0_{gs}$ spectra  with a $J=2$ first excited state was also included. They accounted for a significant fraction $f_{0-2}$ of the total.
\subsubsection{Rotational bands}
The signatures of rotational bands are in both spectral quantities and probes of the wavefunctions. One is therefore concerned with the  energy ratios $R_{82},\,R_{62}$ and $R_{42}$, where $R_{I2}=(E(I_1)-E(0_1))/(E(2_1)-E(0_1)$. Also needed are the reduced transition ratios $B(E2,4_1\rightarrow 2_1)$ and $B(E2,2_1\rightarrow 0_1)$, as well as their ratio $B_{42}$. Values of $R_{42}$ and $R_{62}$ of 10/3 and 7 would be expected for the rigid rotor where $E(J_1)-E(0_{\textrm{gs}}) \sim J(J+1)$.   Pairing has ratios $R_{42}$ and $B_{42}$ close to 0, while vibrational modes have them both close to 2. If the system behaves like a spinning deformed object, these rotational modes give  $R_{42}=10/3$ and $B_{42}=10/7$. The systematics for $R_{42}$ are striking.  The $N=8$, 10, and 12 systems have a sharp peak at $R_{42}=3.3$. This comes solely from the $0-2$ and $J_{\textrm{gs}}=J_{\textrm{max}}$ subsets. The $0_{gs}$ set have an additional peak at 0.5 for N=8, at 0.3 for $N=10$, and at 0.05 at $N=12$. Furthermore, the subset of spectra with $J_{\textrm{gs}}\neq 0$ or $J_{\textrm{max}}$ has a broad peak at $R_{42}=1.4$. The $N=6$ systems has a small peak at 3.3 and a broad peak at 1.3. It is remarkable that in all 6 systems these quantities have narrow peaks at their rotational values for the 0-2 subsets. In \cite{bijker00b} Bijker et al show that the band structure grows with $N/\kappa$ where $\kappa$ is the rank of the interaction, here it is 2. There are some other peaks present for the whole ensemble. In Fig.~\ref{fig.R42B42} we see the results for some 1-level systems.

\subsubsection{Fractional collectivity and the Alaga ratio}
\begin{figure}[htbp]
  \includegraphics[width=\linewidth]{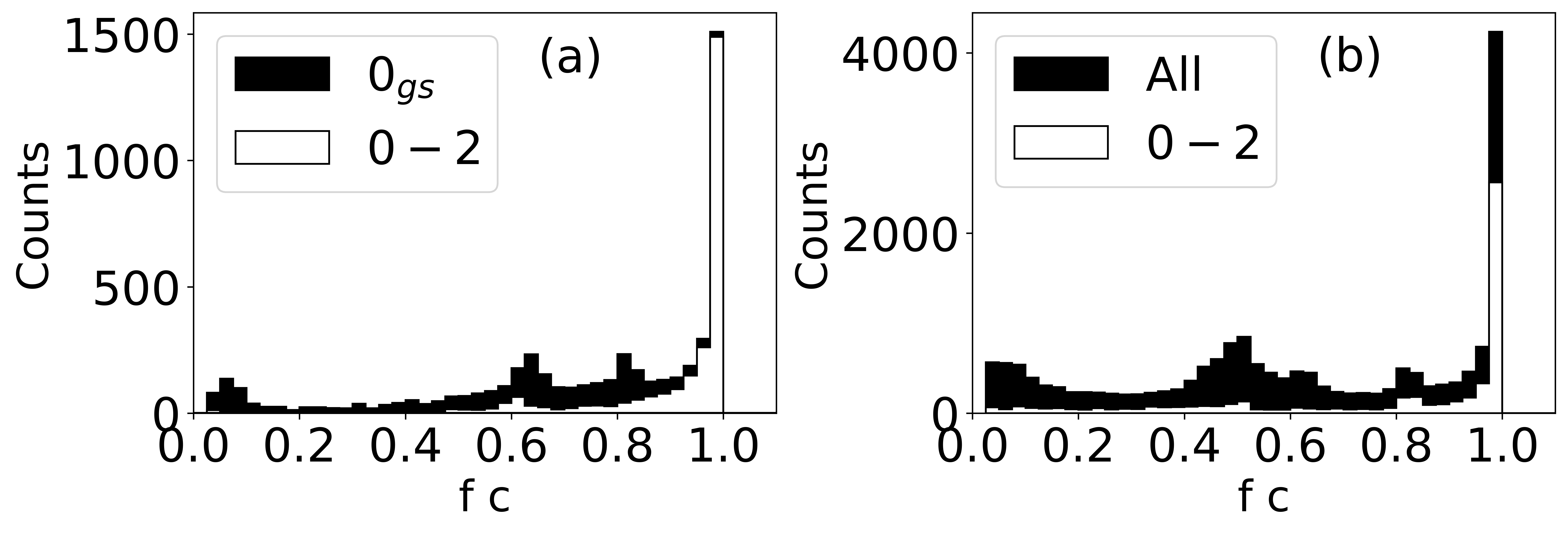}
   \caption{ $(N,j)=(8,4)$. The fractional collectivity $fc$. a) The $0_{\textrm{gs}}$ set is in black. The $0-2$ subset (white) makes up most of the peak at 1.  b) The full ensemble distribution (black) with the $J_{\textrm{gs}}=J_{\textrm{max}}$ subset (white). }\label{fig.fc}
\end{figure}

In their original paper \cite{jbd1} Johnson et al. describe fractional collectivity $fc$ in the $0_{\textrm{gs}}$ wavefunctions of random fermionic systems. The fractional collectivity \cite{jbd1,zelevinsky04} is  defined as
\begin{equation}
    fc=\frac{B(E2,0_1\rightarrow 2_1)}{\Sigma_n B(E2,0_1\rightarrow 2_n)}.
\end{equation}
If the $2_1$ state is made by exciting phonons on the ground state, then $fc$ is close to 1. If the ground state is not collective in nature, then $fc$ is small, and $0_{\textrm{gs}}$ is connected to $2_1$ by manybody operators. $fc$ was evaluated for all $0_1$ states. The results  are striking, see Fig. \ref{fig.fc}. In all 1-level ensembles we see that the $0_1$ state is collective when it is part of the $0-2$ ground state sequence, or when $J_{\textrm{gs}}=J_{\textrm{max}}$. This fact, and the observation that the yrast lines at low-$J$ have similar structures for both these subsets, suggests that the low-$J$ many particle states have the same structure. This is reasonable as they are made with the same $V_L$, the high $L$ values not being relevant to low $J$ states.  Other spurious features were observed, like peaks in $fc$ at 0.23 and 0.63 for $(N,j)=(12,3)$

\begin{figure}[htbp]
  \includegraphics[width=\linewidth]{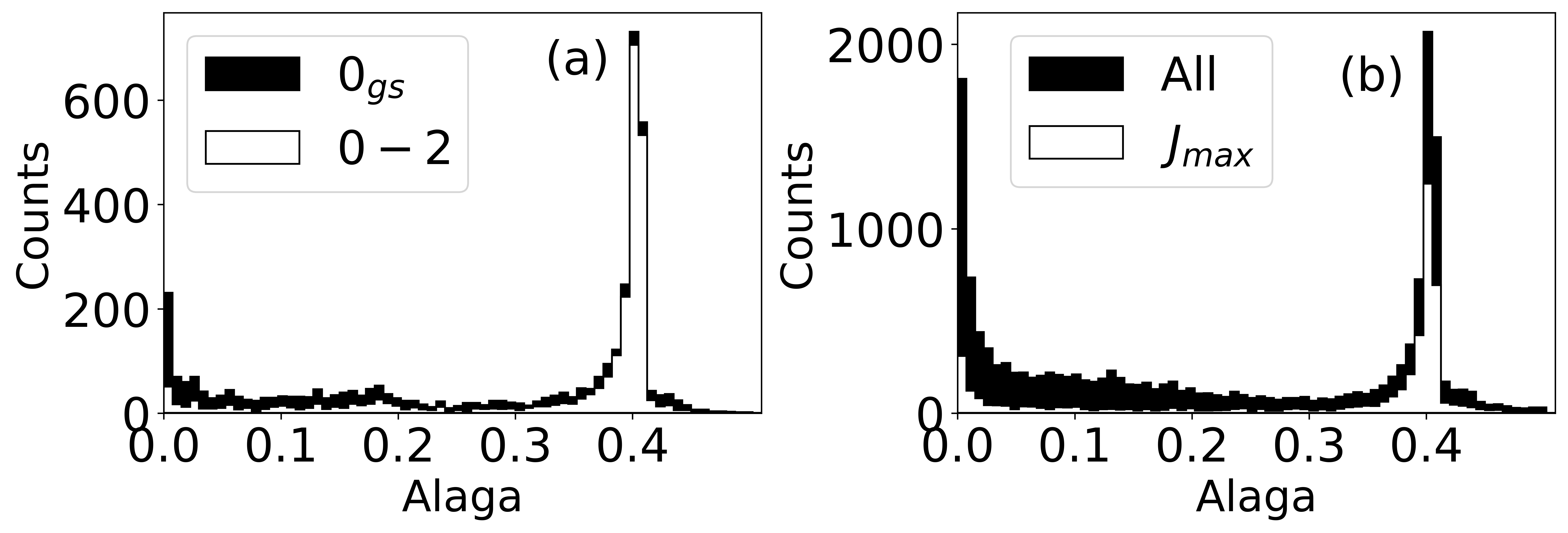}
   \caption{ $(N,j)=(8,4)$. Similar to Fig.~\ref{fig.fc} but the Alaga ratio $A$ is shown.  The results are almost identical.}\label{fig.Alaga}
\end{figure}

\begin{figure}[htbp]
  \centering
  \includegraphics[width=\linewidth]{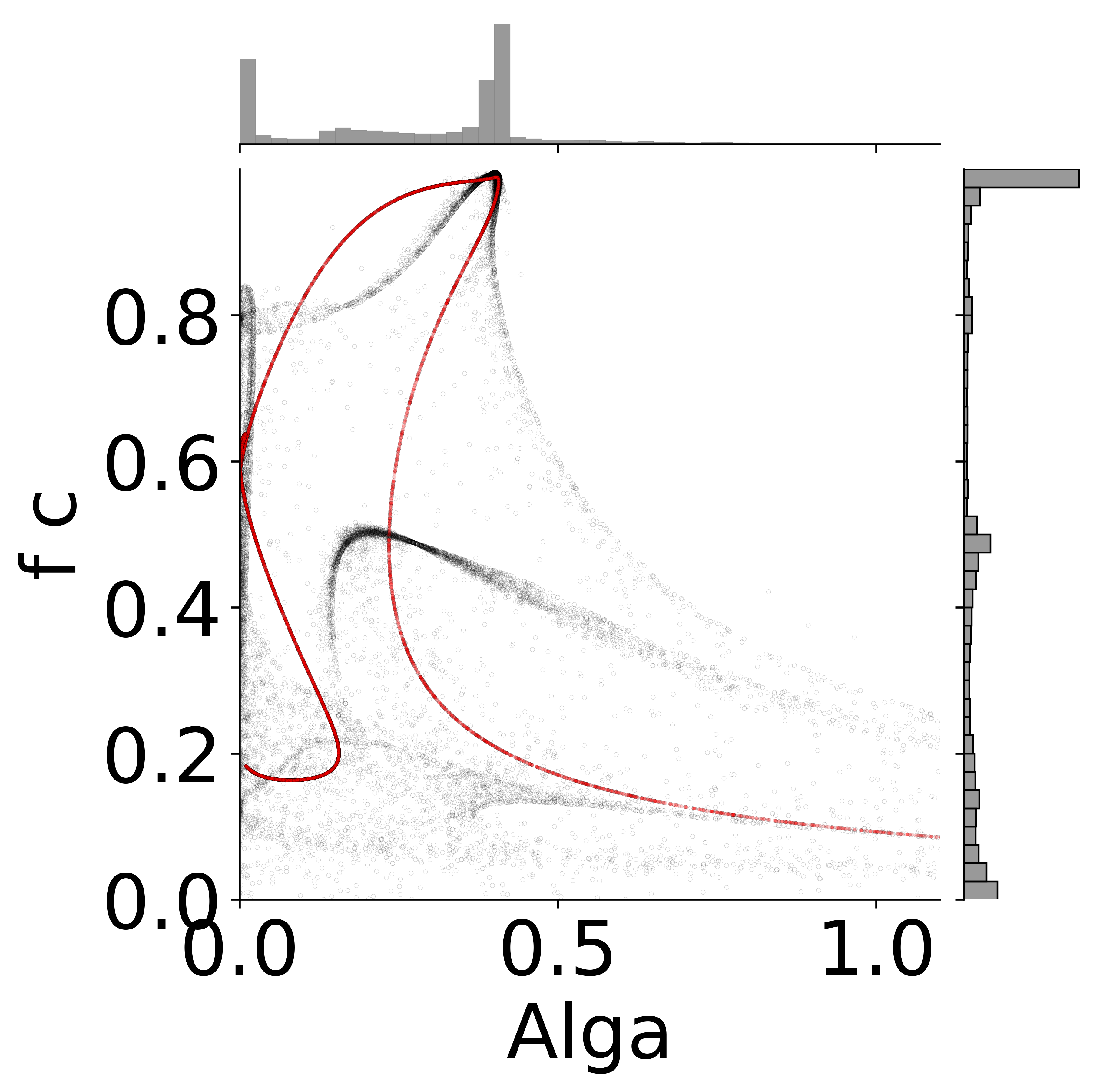}
   \caption{(color online) The Alaga ratio $(A)$ is correlated with the fractional collectivity $(fc)$. Scatterplot of $A$ vs $fc$ for $(N,j)=(10,4)$ (black) and$(12,3)$ (red). There is an obvious strong correlation. The histograms on the axes show the (10,4) data only. The (12,3) data (red) lie  on a smooth line, with no spread.}\label{fig.fcA}
\end{figure}

The Alaga ratio $A=Q^2/B(E2)$  is calculated for $0_1$ states. It is another measure that can reveal collective structure in the wave functions \cite{zelevinsky04,horoi10}.  A nonzero value a corresponds to a rigid rotor, and a zero value corresponds to pairing \cite{bm}. It is closely related to the fractional collectivity so it is not surprising that exactly the same trends appear in $A$ for the 1-level system. There are large sharp peaks at $A=0.43$ in the two cases $0-2$ and  $J_{\textrm{gs}}=J_{\textrm{max}}$.    See Fig.~\ref{fig.Alaga}. The Alaga ratio is correlated with $fc$ in a curious way in the $j=3$ systems. In Fig.~\ref{fig.fcA} we see a scatterplot of $A$ vs $fc$ for $(N,j)=(12,3)$ and (10,4). The (12,3) points lie on a smooth line with no leaking. The $(10,4)$  case is plotted also, with histograms on the margins. There is still a clear strong correlation, but the points spread into the plane in this case.

\subsubsection{Triaxiality}
\begin{figure}[htbp]
  \centering
  \includegraphics[width=\linewidth]{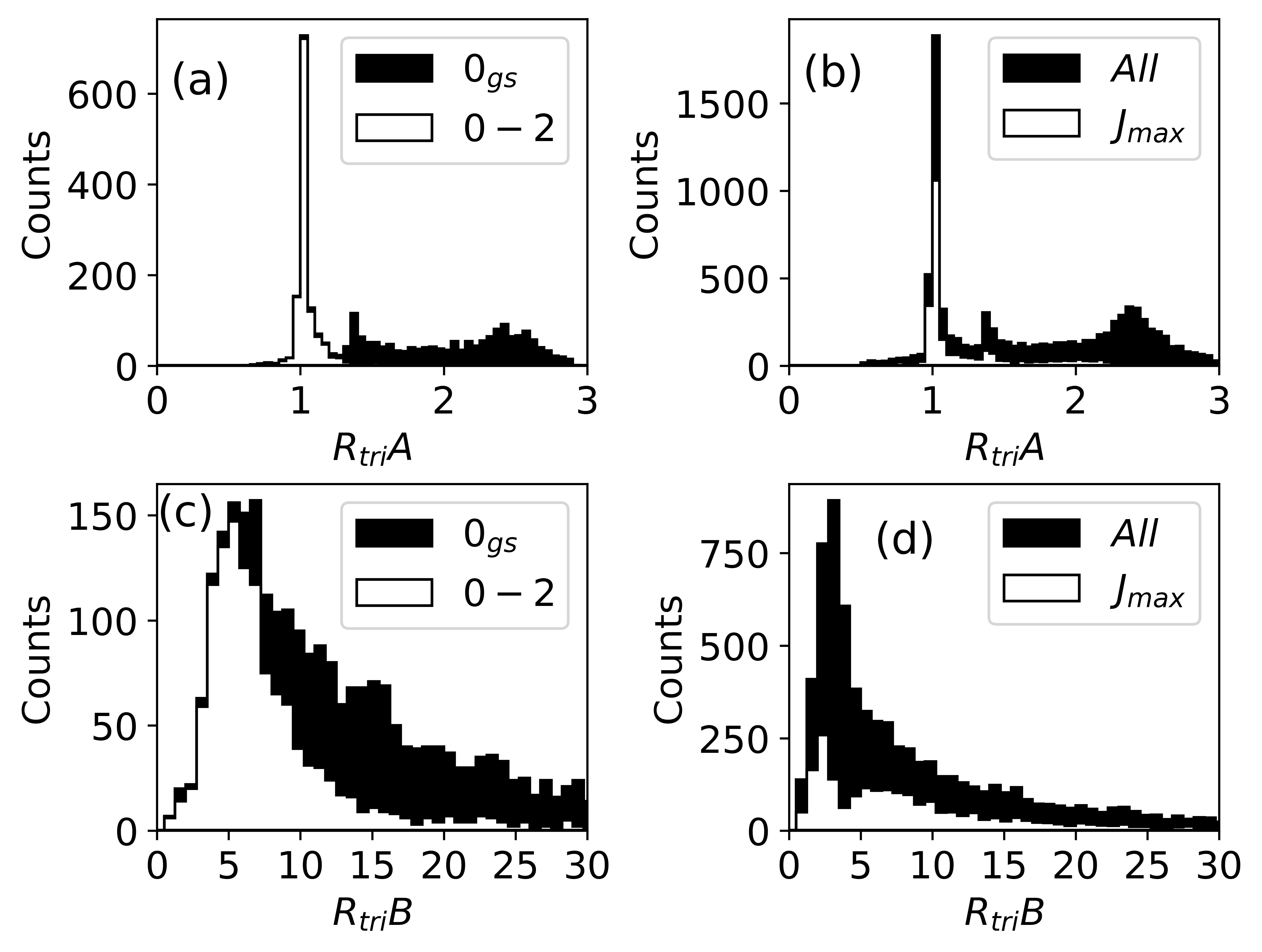}
   \caption{$(N,j)=(8,4)$.(a) $R_{\textrm{triA}}$. The  $0_{\textrm{gs}}$ set (black). The $0-2$ subset (white) makes up most of the peak at 1.  (b) The full ensemble distribution (black) with the $J_{\textrm{gs}}=J_{\textrm{max}}$ subset (white). (c) and (d) Similar to (a) and (b) but with  $R_{\textrm{triB}}$.}\label{fig.tri}
\end{figure}

\begin{figure}[htbp]
  \centering
  \includegraphics[width=\linewidth]{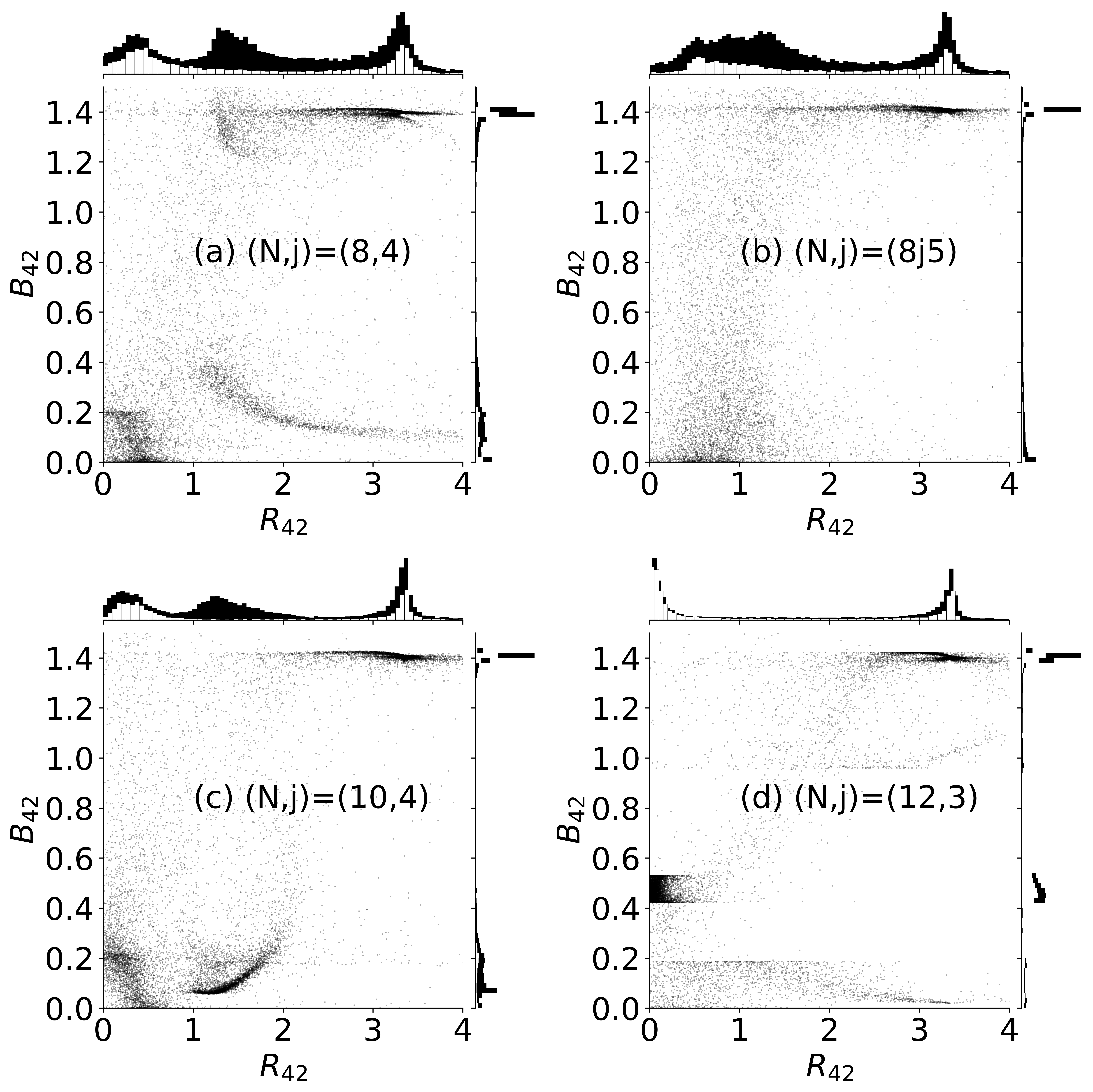}
   \caption{ The scatterplots of $R_{42}$ vs $B_{42}$ for some 1-level systems. In (a) through (d) we the the data clustered about the rotational limit. Also the low-$B_{42}$ values are clumped with the low $R_{42}$ values.}\label{fig.R42B42}
\end{figure}

Low lying $2_2,\, 2_3,\,4_2$ and $5_1$ levels can indicate triaxiality \cite{abramkina11}. The signature is $E(2_1) + E(2_2) = E(3_1)$ and $4 E(2_1) + E(2_2) = E(5_1)$. The quantities $R_{\textrm{triA}}=R_{2_1 3_1 }+R_{2_2 3_1 }$, and    $R_{\textrm{triA}} = 4 R_{ 2_1 5_1} + R_{ 2_2 5_1}$ are shown in Figs.~\ref{fig.tri} for the $(10,4)$ system. As a reminder, $R_{2_1 5_1 } = E(2_1)/E(5_1)$ etc. It is remarkable that $R_{\textrm{triA}}$ has a prominent peak at 1.02. It was observed that as $N$ increased to 12 other peak at 2.75 developed. The story is different for $R_{\textrm{triB}}$, where the strongest feature is a peak at 2 for the $40_{\textrm{gs}}$ spectra, and a peak at 5 for the $0_{\textrm{gs}}$ spectra. These peaks at $R_{\textrm{triA}}=1$   and $R_{\textrm{triB}}=5$ are robust, appearing for $(N,j)=(6,6),\,(8,4),\,(8,5),\,(10,4)$ and (12,3).

\subsubsection{Correlations in quadrupole moments}
\begin{figure}[htbp]
  \centering
  \includegraphics[width=\linewidth]{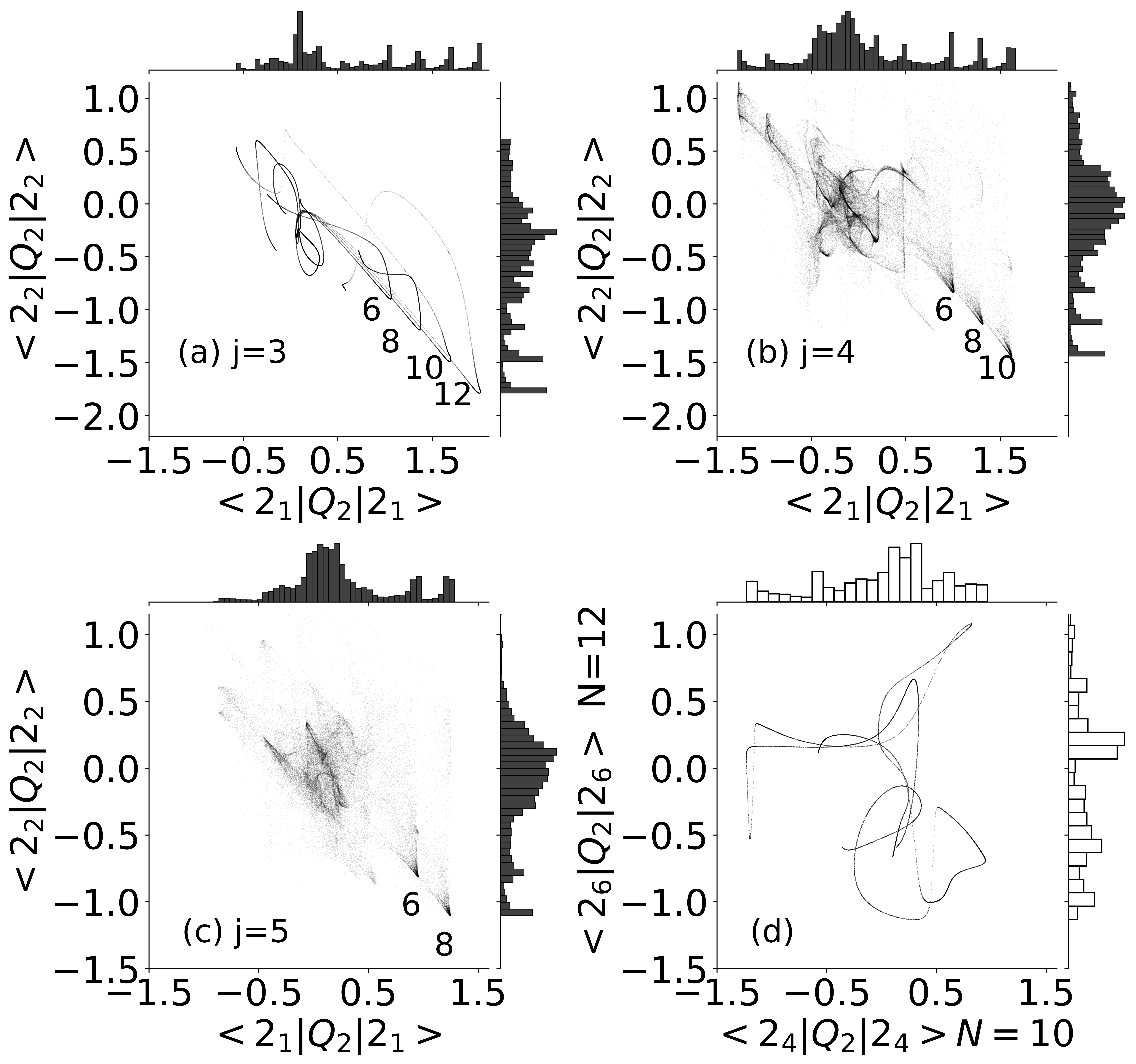}
    \caption{ (a)-(c)  $\mathcal{Q}(2_1)$ vs $\mathcal{Q}(2_2)$. In each case the single particle spin $j$ is fixed. The data for different $N$ is superimposed. The peaks along the line $\mathcal{Q}(2_2) \approx - \mathcal{Q}(2_1)$ are labeled by $N$. Notice in (a) that  data lie along curves. This is a feature of the $(N,j=3)$ space. In (d) we see $\mathcal{Q}$ values for $2_4$, the fourth $J=2$ state in the (10,3) system vs $2_6$ for the (12,3) system lie along a curve.}
    \label{fig:Q2}
\end{figure}

An experimental survey by Allmond \cite{allmond13} revealed correlations in experimental values of $\mathcal{Q}(2_1)$ and $\mathcal{Q}(2_2)$.  The quadropole moment of   the $\textrm{n}^{\textrm{th}}$ state with angular momentum $J$, is $ \mathcal{Q}(J_n)$ defined in Eq.\ref{beq}. A scatterplot places them on a line of slope -1, see Fig.1 in \cite{lei16}. An investigation by Lei \cite{lei16} reproduced these correlations with the two body  random matrix ensemble (TBRE)   in the shell model and   IBM spaces. The 1-level toy model of this work revealed a very strong correlation in the  $0_{gs}$ systems.  Fig~\ref{fig:Q2} shows  scatterplots of $\mathcal{Q}(2_1)$ and $\mathcal{Q}(2_2)$ and see strong and consistent correlations for  $0_{gs}$, with $\mathcal{Q}(2_1) \approx -\mathcal{Q}(2_2)$. Note the histograms on the axes are of the scatterplot show 

The $j=3$ systems  held a surprise. The points lie along lines, and are not at all spread out in the plane.  What is even more surprising is that a scatterplot of $\mathcal{Q}(2_i)$ vs  $\mathcal{Q}(2_j)$ where $2_i$ is and $2_j$ are states from systems with different values of $N$, and $i$ and $k$ have ANY allowed value. To be clear, the $\mathcal{Q}(2_4)$ taken from the $(N,j)=(10,3)$ system plotted against $\mathcal{Q}(2_6)$ from the $(12,3)$ system lie on a line! See Fig.~\ref{fig:Q2}(d). It is tempting to conclude that this correlation,$\mathcal{Q}(2_1) \approx -\mathcal{Q}(2_2)$,  between low lying $J=2$ states is a general feature of rotationally invariant Hamiltonians.

\section{Correlated Yrast Energies \label{2level}}
Rotational and vibrational bands are features of the low energy spectra of many nuclei. They are apparent from the ratios of yrast energies, and the ratios of transition strengths. It came as a surprise when random interactions in the IBM and the vibron model \cite{bijker99, *bijker00a, *bijker02b} showed strong band structure.   The values of $(R_{42},\,R_{62},\,R_{82})$ have specific values depending on the type of collectivity organizing the spectrum. In the IBM, the spectra can be organized by seniority, $S$, the number of unpaired nucleons. This leads to degenerate energies, and the ratios are (1,1,1).  Vibrational spectra have $E(J_1) \propto J$ giving ratios (2,3,4), the $V$ limit. Finally rotational spectra have $E(J_1)\propto J(J+1)$, making the $R$ limit (10/3,7,12). Each of these limits corresponds to a dynamical symmetry of the Hamiltonian, SU(3) for rotational, U(5) for vibrational, and O(5) for seniority. Correlations in yrast energies were examined in \cite{johnson07}. It was seen that a scatterplots of $R_{62}$ vs $R_{42}$ and $R_{82}$ vs $R_{42}$ for 166 stable or ling lived isotopes lay on a trendline containing these three limits. Random interactions on the IBM also showed strong correlations. This "new puzzle" of correlated yrast energies was attacked in the context of the IBM. Lei et al. \cite{lei11} identified and derived new correlations based on different geometric models. They labeled these correlations $\alpha,\,\beta$ and $\gamma$. The correlations are listed in Eq. \ref{corr}, for a full context see \cite{Zhao2018}. $\alpha$ was derived from the anharmonic vibrator model. $\beta$ and $\gamma$ were derived  from the U(5) limit of the IBM, a $d-$boson condensate. Lu et al. \cite{Lu14} identify two more correlations, one of which, $\delta$, is from a $d-$boson condensate. They also saw another correlation around the rotational limit, but it was not well defined in their data. It is well defined in this work, and is labeled   $\rho$.  The set of random parameters $V_L$ in the Hamiltonian populates areas in the parameter space corresponding to the underlying dynamical symmetries of the Hamiltonian.
Wavefunctions corresponding to points along the correlation lines were identified with the different symmetry limits of the IBM by Fu et al. \cite{Fu18}.  This reinforced the physical picture that the $\beta,\,\gamma$ and $\delta$ correlations are from a condensate of $d$ bosons, while the $\alpha$ correlation is a condensate of $s$-bosons.

 \begin{eqnarray}\label{corr}
    \alpha:  R_{62} &=& 3\, R_{42} -3 \qquad \qquad R_{82} = 6 \,R_{42} -8  \\
    \beta:  R_{62} &=& \frac{21}{10}\, R_{42} +0 \qquad \quad\, R_{82} = \frac{18}{5} \,R_{42} + 0   \nonumber\\
    \gamma: R_{62} &=& \frac{9}{5} \,R_{42} +1 \qquad \quad \quad R_{82} = \frac{138}{35} \,R_{42} -\frac{8}{7} \nonumber\\
    \delta: R_{62} &=& \frac{18}{7} \,R_{42}  -\frac{11}{7} \qquad \quad  R_{82} = \frac{33}{7} \,R_{42} -\frac{26}{7} \nonumber\\
    \rho:   R_{62} &=& \frac{21}{4}\, R_{42} -\frac{21}{2}\qquad \quad R_{82} = 18 \,R_{42} - 48  \nonumber\\
\nonumber
 \end{eqnarray}

 \begin{figure}[htbp]
  \includegraphics[width=\linewidth]{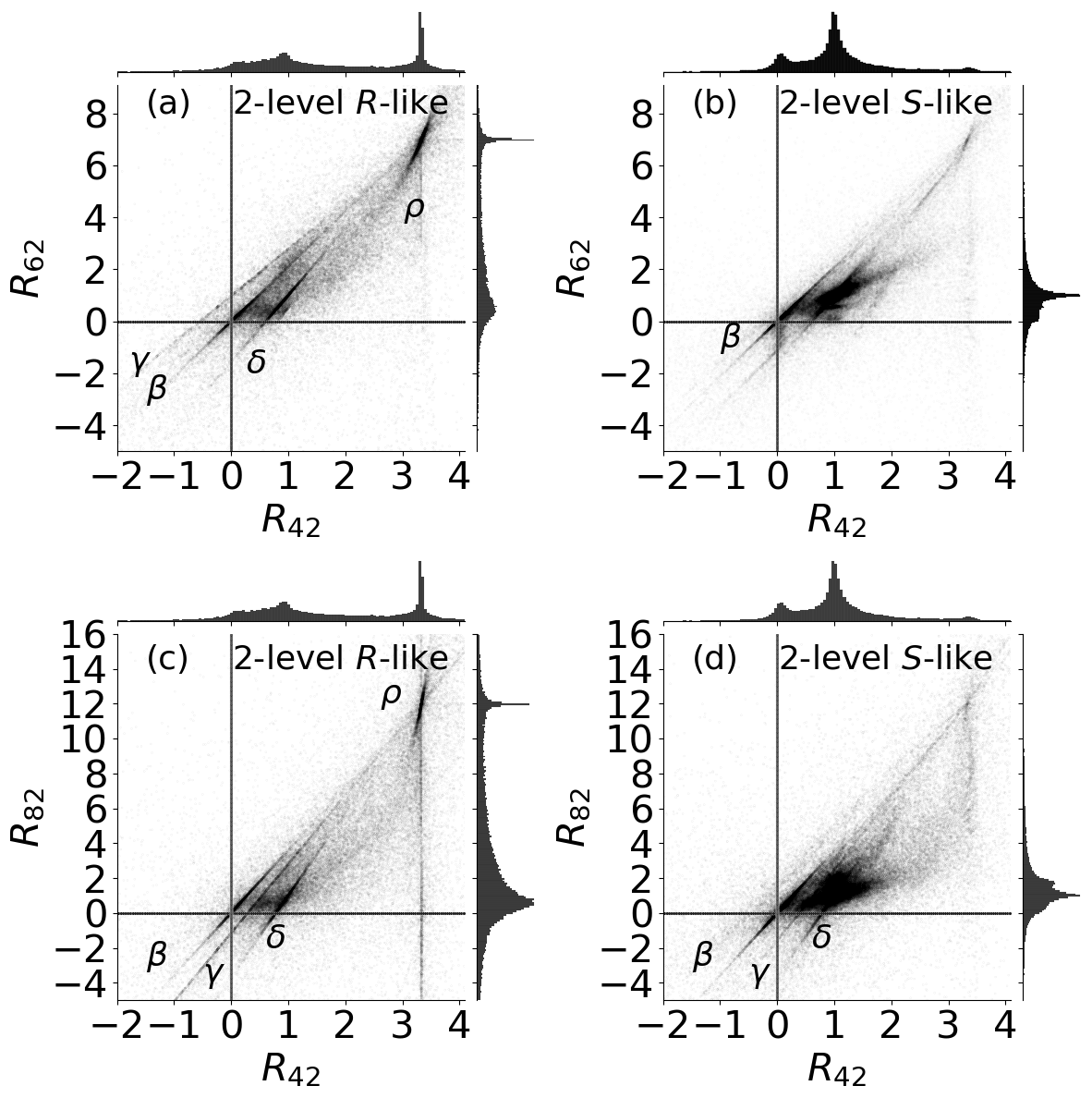}
    \caption{The correlations of the energy ratios $R_{42}$ with $R_{62}$ and $R_{82}$ for $R$-like and $S$-like  groups described in the main text. On the axes are histograms of the $R_{I2}$. Note the peaks at the $S$ and $R$ limits.  (a) $R_{42}$ vs $R_{62}$ for the $R-$like set. The labeled $\beta,\,\gamma,\,\delta$ and $\rho$ correlations are clearly visible. (b) Similar to (a) but for the $S$ group. The $\beta$ and $\delta$ correlations are present but less populated.  (c) and (d) Similar to (a) and (b) but with $R_{82}$. }
    \label{fig:2levelcorr}
\end{figure}

This work   analyzed the yrast energies of  a broad range of random 1 and 2-level systems, not just the IBM. Surprisingly, those same correlations, and some novel ones, are very pronounced in these systems. The 2-level systems are addressed first. A range of $(N,j_1,j_2)$ systems, including  $(N,0,1)$ , and $(N,0,2)$, for $4\leq N \leq 20$, were examined. The results were consistent with previous studies in the vibron model and the IBM \cite{bijker99, *bijker00a, *bijker02b}, so they we will not be rehashed  here. In the other systems the $R_{42}$ distributions had peaks at or close to 0, 1/2, 1 and 10/3. There were also peaks in the transition strength ratio $B_{42}$ at 0, 2 and 2.5. So immediately it is clear  that rotational and vibrational bands are quiet common.  See table \ref{tab:table1}. This made it sensible to group the systems for analysis. The systems with pronounced peaks in $R_{42}$ at 1/2 or 1 were grouped into a set labeled $S$-like, for seniority. This set had $(N,j_1,j_2)$ = (7, 1, 2), (9, 1, 2), (6, 0, 5), (7, 0, 5), (6, 0, 6),(7, 1, 3), (6, 0, 4) ,(7, 0, 4) ,(8, 0, 4) , (6, 2, 3) (6, 3, 4) and  ($N$, 0, 3) with $N = 6, 7\dots 11$. Those systems with a peak in $R_{42}$ at 10/3 were grouped into a set labeled $R$-like, for rotation. This set had $(N,j_1,j_2)$ = (6, 1,2), (8, 1, 2), (10, 1, 2), (6, 1, 3), (8, 1, 3), (6, 1, 4) and (6, 1,5). Each ensemble had 10,000 spectra. Fig.~\ref{fig:2levelcorr}  shows the correlations.

 \begin{figure}[htbp]
  \includegraphics[width=\linewidth]{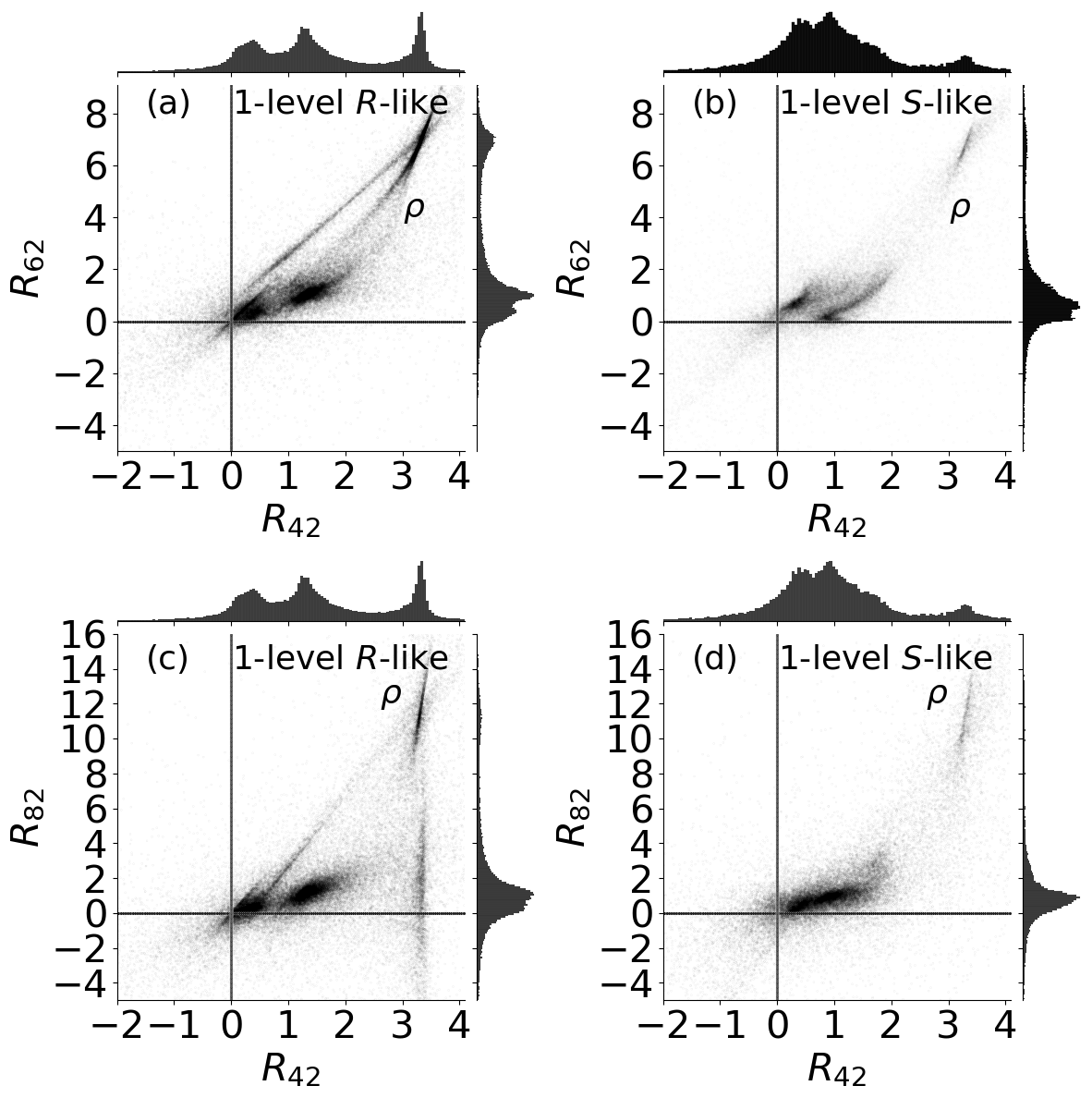}
    \caption{Similar to Fig.~\ref{fig:2levelcorr} but for the 1-level systems. (a) The $\gamma$ and $\rho$ correlations in the $R$-like set stand out. (b) The $\beta$ correlation is clumped about (0.5,1). (c) Compared to the 2-level systems the data are  more disperse. (d) The $\rho$ correlation is faint, and the others are not present. }
    \label{fig:1levelcorr}
\end{figure}

\begin{table}[]
\caption{\label{tab:table1}
The values of $f_0,\,f_{0-2}$, $f_{Jmax}$, the energy ratios $R_{42}$, and the transition strength ratios $B_{42}$, for the $(N,j_1,j_2)$ systems. The  the rotational and seniority limits are common.
}
\begin{ruledtabular}
\begin{tabular}{lllllll}
$(N,j_1,j_2)$&
$f_0$&
$f_{0-2}$&
$f_{Jmax}$&
$R_{42}$&
$B_{42}(\chi=0,1)$\\
\colrule
20,0,1 & 0.68 & 0.13 &   0.27 & 3.3 & 2.52\\
21,0,1 & 0.55 & 0.00 &   0.27 & 3.3 & 2.52\\
18,0,2 & 0.7 \footnote{Period four cycle of $f_0$ vs $N$, see Fig.~\ref{fig.pjd} .} & 0.53 &
0.24&0\footnote{$J_{\textrm{gs}}\neq 0$.}, 2, 3.3 & 0,1.3,1.9,(2.6,2.9)\footnote{Pairs in braces
correspond to $\chi=(0,1)$ in Eq.\ref{m2}}\\
19,0,2 & 0.53 & 0.53 &  0.24&1\footnote{only when $J_{\textrm{gs}}\neq 0$.}, 2, 3.3 & 0,(2.6,2.9)\\
10,0,3 & 0.64 & 0.11 &  0.22& 0,1,3.3 & 0,1.0,2.5\\
11,0,3 & 0.44 & 0.02 &   0.22& 0,1 & 0,1\\
7,0,4 & 0.45 & 0.03 &   0.20 & 0.6 & 0\\
8,0,4 & 0.52 & 0.08 &  0.20& 0.6,3.3\footnote{Small peak.} & 0,2.5\\
6,0,5 & 0.55 & 0.07 &   0.19& 0.9,3.3\footnote{Small peak.} & 0,2.5\\
9,1,2 & 0.17 & 0.01 &   0.21 & - &0, 2.44, 4.7\\
10,1,2 & 0.20 & 0.14 &   0.21 & 0.9\footnote{$J_{\textrm{gs}} \neq 0$.},3.3 & 0, 0.5, 2.4\\
7,1,3 & 0 & 0 &  0.03 &  1 &0\\
8,1,3  & 0.42 & 0.25 &   0.04 &  3.3 & 0, 0.9, 2.3,2.5\\
6,1,4  & 0.33 & 0.15 &   0.18 & 1\footnote{only when $J_{\textrm{gs}} \neq 0$.}, 3.3 & 0,  2.1\\
6,1,5  & 0.38 & 0.16 &   0.17 & 0.5, 3.3 & 0,  2.1, 2.5\\
6,2,3  & 0.38 & 0.16 &   0.17 & 0 , 3.3\footnote{Small peak.} & 0,  2.3, 3($\chi=0$)\\
6,2,4  & 0.41 & 0.09 &   0.15 &   & 0,  2.4\\

\end{tabular}
\end{ruledtabular}
\end{table}

\begin{figure}[htbp]
  \includegraphics[width=\linewidth]{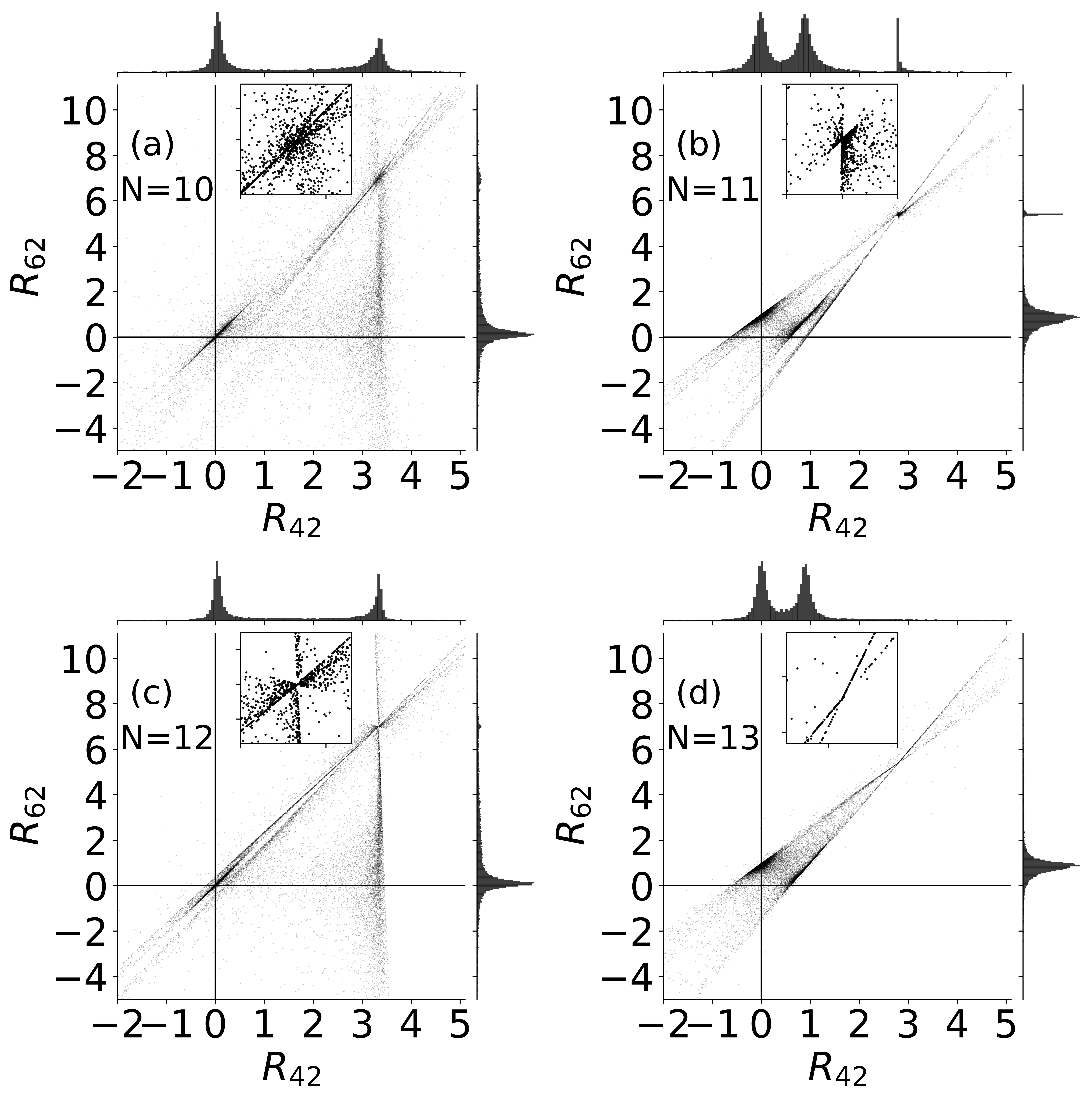}
    \caption{The correlations of the energy ratios $R_{42}$ with $R_{62}$ for $j=3$ with $N$=10 to 13. The $\beta$ correlation is evident in the even-$N$ data. The insets show the zones around the intersection points. These are the rotational limit, (10/3,7) for even-$N$, and (10/3, 5.4) for odd-$N$. }
    \label{fig:j3corr62}
\end{figure}

The 1-level systems were similarly grouped. with the $R$-like set containing $(N,j)$ = (8,4), (8,5) and (10,4). The $S$-like group had $(N,j)$ = (6,6) and (6,7). Each ensemble had 20000. The results are in Fig.~\ref{fig:1levelcorr}. The $N=10,11,12$ and 13 on $j=3$ systems were quiet different from the others. The results are plotted separately. There are two interesting points that stand out immediately. The first is the very clear linear correlations that separate the space in the odd-$N$ systems. The other thing is the structure around the points of intersection of the linear correlation lines. They have very clear geometrical exclusion zones. This is shown in the insets, see Fig.~\ref{fig:j3corr62} and \ref{fig:j3corr82}. The rotational limits were clear on the even-$N$ $j=3$ systems, but for odd-$N$, the lines intersected the points (3.3,5.4) in $R_{42}$ vs $R_{62}$ and (3.3,8.8) in $R_{42}$ vs $R_{82}$. No attempt was made to derive the correlations other than to point out that $\beta$ is still prominent in the even-$N$ systems.

The $\beta,\,\gamma$ and $\delta$ correlations are clearly present in the 2-level systems, while shades of them are seen in the 1-level systems.  The $\rho$ correlation is seen clearly in both 1 and 2-level systems. This raises questions as to the origin and interpretation of these correlations. If the results in Eq.\ref{corr} are contingent on the models used, (random IBM, shell model, fermion dynamical symmetry model etc) then why do they work on such a wide range of systems. Is it still reasonable to regard these correlations as condensates of $d-$boson clusters, if they happen in systems with $(j_1,j_2)=(1,3),\,(1,4)$ and (1,5), or even 1-level systems?

\section{Conclusions}
\begin{figure}[htbp]
  \includegraphics[width=\linewidth]{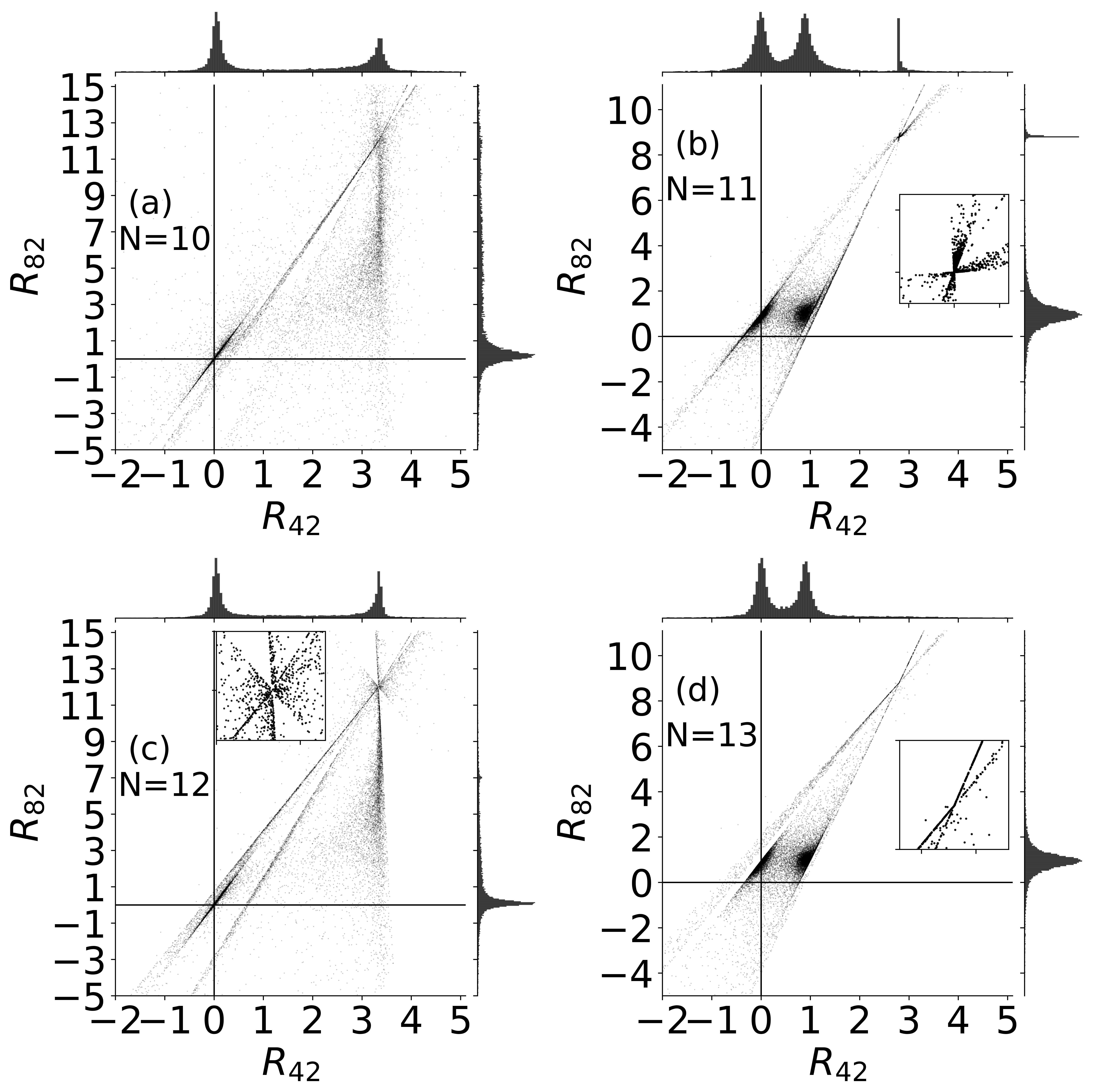}
    \caption{Similar to Fig. \ref{fig:j3corr62} but here we plot  energy ratios $R_{42}$ with $R_{82}$. }
    \label{fig:j3corr82}
\end{figure}

Random interactions on boson systems continue to surprise. The yrast lines for 1-level systems reveal robust structures that depend of the  ground state spin. The $J_{\textrm{gs}}=J_{\textrm{max}}$ spectra had a stepped structure that could be understood as a condensate of non interacting spin-$2j$ pairs. The $J_{\textrm{gs}}=J_{Nj-N}$ spectra in 1 and 2-level systems had yrast lines that lay on a common line upon rescaling. This is interpreted as a rotational band based on a deformed rotating hard core ground state. When the ground state is 0($j$) for even (odd) $N$ systems, the yrast lines are a series of kinked parabolas, with kinks separated by $2j$. These kinks are accompanied by jumps in the dynamical moment of inertia. The 1-level systems had pronounced peaks in fractional collectivity ($fc)$ and the related Alaga ratio. In both cases the indication was that the $0_1$ and $2_1$ states are very collective when the ground state sequence is 0-2, or when $J_{\textrm{gs}}=J_{\textrm{max}}$. These same sets of spectra exhibited a sharp peak in $R_{\textrm{triA}}$. The random 1-level systems continued to behave like real world spectra with $\mathcal{Q}(2_1) \approx -\mathcal{Q}(2_2)$, which is what happens in experimental nuclear data also. The low yrast energies of the 2-level systems were highly correlated, mimicking the case of the symmetry limits if the IBM. The systems involved had 6 or more particles on $(j_1,j_2)$ quiet different from (0,2). Surprisingly the 1-level systems also showed these same strong correlations. The $j=3$ systems had their own special behavior, quite surprising at times. New and strong energy ratio correlations that were previously understood in the random IBM as condensates of $s$ and $d$ bosons have reappeared in a different context. They are in 1 and 2-level systems with a range of single-particle spin. Excluded regions in these energy correlations appear when $j=3$ (see insets of Figs.\ref{fig:j3corr62},\ref{fig:j3corr82}). Further geometric puzzles appear in the $j=3$ systems where the quadrupole moments are correlated and scatterplots of $\mathcal{Q}(2_1)$ vs $\mathcal{Q}(2_2)$ lie on lines, as does the scatterplots of fractional collectivity and the Alaga ratio. All these observations are an invitation to further study, and an indication of how rich these simple systems continue to be.

\begin{acknowledgments}
I wish to acknowledge the support of the Office of Research Services of the University of Scranton. Some of this work was done while visiting Instituto de Ciencias Nucleares, Universidad Nacional Aut\'{o}noma de M\'{e}xico, and the Physics department at University College Dublin. I am grateful to R. Bijker, A. Volya , and V. Zelevinsky for constructive discussions.
\end{acknowledgments}

\bibliography{orderedbosonsystems}% Produces the bibliography via BibTeX.

\end{document}